\documentclass[12pt,draftcls,onecolumn]{IEEEtran}

\usepackage[noadjust]{cite}

\usepackage[table]{xcolor}
\usepackage{diagbox}
\usepackage[pdftex]{graphicx}
\usepackage{adjustbox}
\usepackage{tikz}
\usepackage{subfig}
\usetikzlibrary{shapes,arrows,automata}
\usepackage[cmex10]{amsmath}
\usepackage{amssymb}
\usepackage{bm}

\usepackage{algorithm}
\usepackage{algorithmic}
\usepackage{textcomp,booktabs}
\usepackage{colortbl}
\definecolor{mygray}{gray}{.9}
\newcolumntype{g}{>{\columncolor{mygray}}c}

\tikzset{
relay node/.style={circle, minimum size=0.75cm, inner sep=0pt, draw=blue!70, fill=blue!20},
end node/.style={circle, minimum size=0.75cm, inner sep=0pt, text width=6mm, align=center, draw=red!80, fill=red!20},
any node/.style={circle, minimum size=0.65cm,draw=mygray,fill=mygray},
}

\usepackage{chngcntr}

\begin{document}
%
\title{Practical Inner Codes for Batched Sparse Codes in Wireless Multi-hop Networks}
%
%
%

\author{Zhiheng~Zhou,~\IEEEmembership{Member,~IEEE,}
        Congduan~Li,~\IEEEmembership{Member,~IEEE,}
        Shenghao~Yang,~\IEEEmembership{Member,~IEEE,}
        and~Xuan~Guang,~\IEEEmembership{Member,~IEEE}
\thanks{Z.~Zhou is with the National Key Laboratory of Science and Technology on Communications, University of Electronic Science and Technology of China, Chengdu, China (e-mail: zhzhou@uestc.edu.cn).}
\thanks{C.~Li is with the Department of Computer Science, City University of Hong Kong, Tat Chee Ave., Hong Kong SAR, China (e-mail: congduan.li@cityu.edu.hk).}
\thanks{S.~Yang is with the School of Science and Engineering, the Chinese University of Hong Kong (Shenzhen), Shenzhen, China (e-mail: shyang@cuhk.edu.cn).}
\thanks{X.~Guang is with the Institute of Network Coding, The Chinese University of Hong Kong, Hong Kong SAR, and the School of Mathematical Sciences, Nankai University, Tianjin, China (e-mail: xuanguang@inc.cuhk.edu.hk).}}


\maketitle

\begin{abstract}
Batched sparse (BATS) code is a promising technology for reliable data transmission in multi-hop wireless networks. As a BATS code consists of an outer code and an inner code that typically is a random linear network code, one main research topic for BATS codes is to design an inner code with good performance in transmission efficiency and complexity. In this paper, this issue is addressed with a focus on the problem of minimizing the total number of packets transmitted by the source and intermediate nodes. Subsequently, the problem is formulated as a mixed integer nonlinear programming (MINLP) problem that is NP-hard in general. By exploiting the properties of inner codes and the incomplete beta function, we construct a nonlinear programming (NLP) problem that gives a valid upper bound on the best performance that can be achieved by any feasible solutions. Moreover, both centralized and decentralized real-time optimization strategies are developed. In particular, the decentralized approach is performed independently by each node to find a feasible solution in linear time with the use of look-up tables. Numerical results show that the gap in performance between our proposed approaches and the upper bound is very small, which demonstrates that all feasible solutions developed in the paper are near-optimal with a guaranteed performance bound.
\end{abstract}


\IEEEpeerreviewmaketitle

\newtheorem{theorem}{Theorem}
\newtheorem{lemma}{Lemma}
\newtheorem{proposition}{Proposition}
\newtheorem{corollary}{Corollary}
\newtheorem{remark}{Remark}
\newtheorem{definition}{Definition}

\section{Introduction}
%
%
%
%
\IEEEPARstart{M}{ulti-hop} wireless networks have many applications, such as the wireless sensor networks, the underwater networks and the vehicular networks. In these scenarios, a wireless device or a source node would like to reliably transmit data to a destination node via multiple intermediate relay nodes. However, severe packet losses may occur in wireless communications due to the multipath effect, congestion, limited resources and hidden nodes. The more the number of hops is, the higher the probability that a packet losses becomes. To provide end-to-end reliability in multi-hop wireless networks, various techniques, such as retransmission~\cite{Fu05}, network coding~\cite{Ahlswede00}-\cite{Chi10} and fountain codes~\cite{Shokrollahi06}\cite{Luby11}, have been proposed to handle the packet loss. However, these mechanisms are not efficient for multi-hop systems and lead to a waste of resources~\cite{YangAllerton14}\cite{Zhang16}.

BATched Sparse (BATS) code \cite{YangTIT14} as a promising new technique, is proposed to achieve the end-to-end reliable communications in multi-hop wireless networks. A BATS code consists of {\it outer code} and an {\it inner code} \cite{YangTIT14}. The outer code is a matrix generalized fountain code to generate a potentially unlimited number of batches. Each batch consists of $M$ coded packets. At each forwarding node (including the source node), a random linear network code as an inner code is used for the packets of the same batch in order to overcome the accumulation of the packet loss over multi-hop transmissions. The inner code directly affects the empirical rank distribution that plays a crucial role for the design of the outer code. The destination node can utilize a belief propagation (BP) decoder to retrieve the source messages from the received batches with low complexity.

BATS codes preserve the salient feature of fountain codes, in particular, the rateless property, which can significantly reduce the number of acknowledgements and avoid retransmission. On the other hand, with the use of relatively small batch size, BATS codes have lower encoding/decoding complexity and less caching requirements in the intermediate nodes, compared with the ordinary random linear network coding schemes~\cite{Li03}. Moreover, BATS codes generally achieve higher rates than some other low-complexity random linear network coding schemes, such as EC codes~\cite{Tang12} and Gamma codes~\cite{Mahdaviani12}.

As a crucial component of BATS code, the design of inner codes has received widely attentions~\cite{Tang16}\cite{Yin16}. Tang \textit{et al.}~\cite{Tang16} pointed out that the random scheduling scheme is not efficient for inner codes in line networks, and instead presented an adaptive scheduling method to optimize the end-to-end throughput. Yin \textit{et al.}~\cite{Yin16} further designed an algorithm, named Adaptive Recoding (AR), where each intermediate node chooses the number of coded packets of a batch in accordance with both its rank and the rank distribution of all received batches. In \cite{Yin16}, the authors also numerically computed the throughput on line networks with the assumption that all nodes can communicate simultaneously.

However, in the context of wireless networks, not all nodes within a certain area (e.g., sender and its neighbors) can transmit data at the same time due to the limited bandwidth resources. In order to efficiently utilize bandwidth resources to support high throughput, the total number of packets transmitted along a flow path should be carefully considered. In addition to bandwidth constraint, wireless nodes are typically powered by batteries that are of limited capacity and even non-replaceable in many applications. One of the biggest consumers of energy is data transmission~\cite{Muruganathan05}\cite{Li10}. Hence, communication cost highly depends on the number of packets sent. Therefore, it is important to optimal the total number of transmitted packets between source and destination to save energy and improve network lifetime.

In this paper, we investigate the optimal inner code problem for BATS code in multi-hop wireless networks, with the objective of minimizing the expected number of transmissions from the source to the destination node. The problem is formulated as a Mixed Integer NonLinear Programming (MINLP) problem, which is typically difficult to solve. To bypass the difficulty, we investigate the inherent structures in the inner codes and develop an upper bound on the optimal solution. In the meantime, the centralized and decentralized real-time optimization approaches are proposed. By using these approaches, the total number of transmissions can be significantly reduced as well as achieving the high transmission efficiency. Our results fill in some important gaps in the current understandings on optimizing the inner codes in wireless networks. Specifically, four major contributions of this paper are as follows.

\begin{itemize}
    \item We establish the relation between the empirical rank distribution and the total number of transmission. By using the recursive expression of the rank distribution, we show that minimizing the total number of transmissions can be formulated as an MINLP problem. To the best of our knowledge, this is the first attempt to investigate the number of transmissions from an end-to-end point of view for BATS codes.

    \item We model the channel as an one-step Markov process. Based on the property of the eigen-decomposition of the transition matrix, the explicit formula for evaluating the rank distribution is derived. Further, by utilizing both the explicit formula and the incomplete beta function \cite{Zelen72}, a NonLinear Programming (NLP) problem is constructed. We, then, prove that this NLP provides a valid upper bound on the optimal value of the MINLP.

    \item We propose both centralized and decentralized real-time optimization stategies for designing the inner codes in multi-hop wireless networks. The centralized scheme is performed by the source only, while the decentralized one with linear complexity is independently operated by each node. More specifically, each node solves the optimization problem by means of the look-up tables that are built from the properties of the objective function and the information obtained from its next hop. In addition, a variation of the look-up table designed to adapt to the dynamic networks, is also discussed.

    \item We show that the proposed approaches yield an objective value very close to the upper bound, which indicates that our approaches can offer the near-optimal solutions. We also evaluate the performance of the proposed approaches on the average rank of the received batches, which provides some guidelines for future algorithm and protocol designs in practical networks.
\end{itemize}

The remainder of the paper is organized as follows. In \S II, we start with a brief introduction to the BATS codes and the related works. In \S III, the network model is discussed with the formulation of the MINLP problem. \S IV  establishes a NLP problem by relaxing integrity constraint of the MINLP. In addition, the practical schemes to find the optimum in centralized and distributed methods are proposed, respectively. In \S V, extensive simulation results are presented to illustrate that the real-time approaches presented in \S IV are able to offer near-optimal solution. \S VI concludes the paper.

\section{Background}
\subsection{Batched Sparse Codes}


A BATS code consists of an outer code and an inner code. The outer code is only performed by source nodes. Suppose a source node needs to send $N$ input packets to a destination node via a wireless network, where each symbol of a packet is an element of the finite field $\mathbb{F}_q$, where $q$ is a prime power. Fix an integer $M \geq 1$ as the batch size. Using the outer code, a sequence of batches $\mathbf{X}_i, i = 1,2 \ldots$ are generated as,
\begin{equation}
\mathbf{X}_i = \mathbf{B}_i \cdot \mathbf{G}_i, \nonumber
\end{equation}

\noindent where $\mathbf{B}_i$ is a matrix consisting of $\text{dg}_i$ columns, each of which is a source packet that is randomly picked out, and $\mathbf{G}_i$ is a totally random matrix on $\mathbb{F}_q$ of size $\text{dg}_i \times M$, i.e., each entry of $\mathbf{G}_i$ is independently and uniformly chosen from $\mathbb{F}_q$. Here, $\text{dg}_i$ is called the \textit{degree} of the $i$-th batch $\mathbf{X}_i$. The degrees $\text{dg}_i, i = 1,2, \ldots ,$ are independent identically distributed (i.i.d.) random variables with a given distribution $\mathbf{\Psi} = (\Psi_1, . . . , \Psi_N)$, i.e., $\Pr \{\text{dg}_i = n \} = \Psi_n, 1 \le n \le N$.

Before transmitting batch $\mathbf{X}_i$, the source node performs random linear network coding on packets belonging to the batch. When an intermediate node receives a batch, the node will also apply random linear network coding to the batch and, then, forward it. The procedure is called inner coding (or recoding).

In particular, the batch $\mathbf{Y}_i$ received by a node can be expressed as
\begin{equation}
\mathbf{Y}_i = \mathbf{X}_i \cdot \mathbf{H}_i, \nonumber
\end{equation}

\noindent where $ \mathbf{H}_i$ is called \textit{transfer matrix}. After the destination node receives enough batches, the source packets can be efficiently decoded by using belief propagation (BP).

\subsection{Related Works}
Ng \textit{et al.} \cite{Ng13} studied the performance of finite-length BATS codes with respect to BP decoding. In \cite{YangAllerton14}, \cite{Huang14} and \cite{Zhang16}, the authors proposed a BATS-based network protocol and evaluated the performance over lossy channels. In particular, Yang \textit{et al.} \cite{YangAllerton14} designed a simple packet interleaving scheme to combat against the bursty losses. In the mean time, Huang \textit{et al.} \cite{Huang14} proposed a FUN framework, where an inner-encoding algorithm was designed to mix the packets belonging to two intersecting flows. Zhang \textit{et al.} \cite{Zhang16} further extended their previous work \cite{Huang14} and indicated both theoretically and practically that their algorithms performed better than the exiting approaches in TDMA multi-hop networks. In \cite{Xu16}, the authors proposed a distributed two-phase cooperative broadcasting protocol, which uses BATS codes in the first phase to help the peer-to-peer (P2P) communications.

\section{System Model and Problem Formulation}
\subsection{Network Model}
We consider a multi-hop wireless network where one source node intends to deliver packets to one sink node. Both source and sink nodes are arbitrarily placed in the network, while the sink node is out of the transmission range of the source. Consequently, the end-to-end transmission needs help from intermediate nodes. The communication process consists of two phases: the initial phase and the transmission phase.

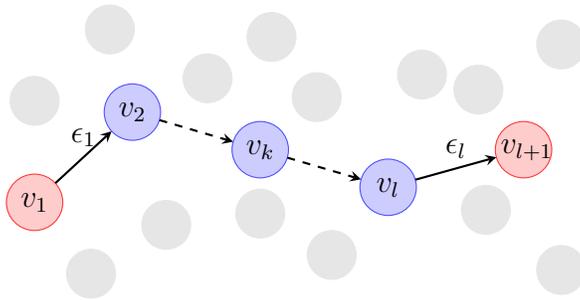
\begin{figure}
\centering
\begin{tikzpicture}[auto, outer sep=0pt, >=stealth]
\foreach \place/\x in {{(-3.5,0.65)/1}, {(-2.75,-1.65)/2},{(-1.75,-1)/3},
    {(-1.2,0.95)/4}, {(-0.35,1.5)/5}, {(-0.5,-0.85)/6}, {(0.25,0.7)/7}, {(0.45,-1.4)/8}, {(1.65,1)/9}, {(2.4,0.8)/10}, {(2.5,-0.8)/11}, {(3.5,1.4)/12}, {(3.5,-1.6)/13}, {(-2.5, 1.6)/14}}
\node [any node] (a\x) at \place {};

\node [end node] (S) at (-3.5,-0.7) {$v_1$};
\node [end node] (D) at (3,0) {$v_{l+1}$};
\node [relay node] (R1) at (-2.2,0.5) {$v_{2}$};
\node [relay node] (Rk) at (-0.5,-0) {$v_{k}$};
\node [relay node] (R2) at (1.2,-0.5) {$v_{l}$};

\draw [->,thick] (S) --  node[above] {$\epsilon_1$} (R1);
\draw [dashed,->,thick] (R1) -- (Rk);
\draw [dashed,->,thick] (Rk) -- (R2);
\draw [->,thick] (R2) --  node[above] {$\epsilon_l$} (D);

\end{tikzpicture}
\caption{The network with an unicast flow.}
\label{fig:1}
\end{figure}

In the initial phase, the source node establishes a path to the sink node by means of a single-path routing protocols (e.g., DSDV \cite{Perkins94}). For instance, the path can be described as Fig. \ref{fig:1}, which consists of $l + 1, l > 0$ nodes $v_k, 1 \le k \le l + 1$. Nodes $v_1$ and $v_{l+1}$ denote the source and sink, respectively. In addition, the source collects the quality of each link, e.g., packet loss rate (PLR). Let $\epsilon_{v_kv_{k+1}}$ (or $\epsilon_k$ for short) denote the PLR from node $v_k$ to node $ v_{k+1} $. In the model, we consider that packet losses are governed by i.i.d. Bernoulli processes.

In the transmission phase, the source generates and sends a set of message packets to the sink node along the path. The source packets are encoded by a BATS code. More specifically, the source performs both the outer and inner encoding, while the intermediate nodes only recode their received packets by means of an inner code. Let $\mathbf{X}_{in}$ be a batch cached by node $v_k, 1 \le k \le l$, then the inner code can be expressed by
\begin{equation}
\label{eq:inner_exp}
\mathbf{X}_{out} = \mathbf{X}_{in} \cdot \mathbf{\Phi},
\end{equation}

\noindent where the elements of matrix $\mathbf{\Phi}$ are independently and randomly chosen from some finite field. We call $\mathbf{\Phi}$ totally random coding matrix (or coding matrix for short). As we will show that the number of columns of $\mathbf{\Phi}$ is closely related to the performance of a BATS code.

In addition, only the end-to-end ACKs are allowed during the transmissions in the last phase. It is postulated that the route remains available until the traffic session is closed.

\subsection{Problem Formulation}
Let $n_k$ be the number of batches transmitted by node $v_{k}, 1 \le k \le l$. Let index set $\Omega_k$ consist of the indices of batches received by node $v_k (1 \le k \le l + 1)$ and, particularly, $\Omega_1 \supseteq \Omega_2 \supseteq \cdots \supseteq \Omega_{l+1}$. Let $t_{k,i}$ be the number of coded packets generated by node $v_{k}$ for the $i$-th batch ($i \in \Omega_k$). In this paper, we let node $v_k$ generate the same number of coded packets, say $t_k$, for every received batch in $\Omega_k$, i.e., $t_{k,i} = t_k$, $\forall~i\in \Omega_k$.

Since every batch sent by node $v_k$ consists of $t_k$ coded packets, the total number $T_{total}$ of packets transmitted by the source node and all the intermediate nodes during the communication is
\begin{equation*}
\label{eq:T_total}
T_{total} = \sum_{k = 1}^{l} n_k t_k.
\end{equation*}

In this paper, our goal is to minimize $T_{total}$ as mentioned before that it is closely related to the wireless network performance such as throughput and energy. Toward this end, we need determine the optimal values of $n_k$ and $t_k$. It is worth to note that each node can only process the batches they received. Since packet losses are i.i.d. as well as batches are generated independently by node $v_k$, the number $n_{k+1}$ of batches received by $v_{k+1}$ can be evaluated to $n_{k+1} = n_k (1 - \epsilon_k^{t_k})$. Thus, we obtain
\begin{equation}
\label{eq:T_total_stat}
T_{total} = n_1 \sum_{k = 1}^{l} \prod_{i=1}^{k} (1 - \epsilon_i^{t_i}) t_k
\end{equation}

After the both sides of Eq. (\ref{eq:T_total_stat}) are normalized by the number of source packet, we define \textit{transmission efficiency} as
\begin{equation*}
\eta = \frac{N}{T_{total}}
=
\frac{N} {n_1} \cdot \frac{1} {\sum_{k = 1}^{l} \prod_{i=1}^{k} (1 - \epsilon_i^{t_i}) t_k}.
\end{equation*}

Suppose that the empirical rank distribution of the transfer matrices converging to $ \mathbf{h}_{k} = [h_{k,0}, h_{k,1}, \ldots, h_{k,M}]$ at node $v_k,  1 \le k \le l+1$. In particular, $ \mathbf{h}_{1} = [0, 0, \ldots, 0,  1]$ at the source node. All vectors are column vectors throughout this paper. Then, we can design an outer code such that the coding rate $N / n_1 \approx \sum_{r=1}^{M} r h_{l+1,r} \triangleq \hbar_{l+1}$ when $N$ is sufficiently large \cite{YangTIT14}. In this paper, we use the average rank $\hbar_{l+1}$ to approximate $N / n_1$ and, then, redefine \textit{transmission efficiency} $\eta$ as
\begin{equation}
\label{eq:batch_transmission_efficiency}
\eta =
\frac{\hbar_{l+1}} {\sum_{k = 1}^{l} {\prod_{i=1}^{k} (1 - \epsilon_i^{t_i}) t_k}}.
\end{equation}

Next, let us show how to estimate $\hbar_{l+1}$ by means of $t_i, 1\le i \le l$. We have the following theorem.

\begin{theorem}\em
\label{th:rank_dist}
Let
\begin{equation}\label{zeta_r_n_m}
\zeta_{r}^{n,m} = \frac{\zeta_{r}^{n} \zeta_{r}^{m}}{\zeta_{r}^{r} q^{(n - r) (m - r)}},
\end{equation}
and
\begin{equation}\label{zeta_r_n}
\zeta_{r}^{n} =
\begin{cases}
\prod\limits_{i=0}^{r-1} (1 - q^{-n+i}), & r > 0; \\
1, & r = 0.
\end{cases}
\end{equation}
Then the probability that the rank of transfer matrices obtained by node $v_{k+1}, k = 1, 2, \ldots, l,$ is $r$ is \begin{align}
\label{eq:rank.dist}
h_{k+1,r} = \sum\limits_{m=r}^{M} {\sum\limits_{n=r}^{t_k} {h_{k,m} \binom{t_{k}}{n} (1 - \epsilon_k)^n \, \epsilon_k^{t_{k} - n} \, \zeta_{r}^{n,m}}},
\end{align}
under the boundary condition $\mathbf{h}_{1} = [0, \cdots 0, 1]$.
\end{theorem}
\begin{IEEEproof}
The proof is given in Appendix~\ref{app-A}.
\end{IEEEproof}

Clearly, the average rank $\hbar_{i}$ is a function of $t_j, j = 1,2, \ldots, i - 1$, with given $M$ and $q$. Now, we construct the optimization problem $\mathbf{P1}$ as follows.
\begin{align*}
\mathbf{P1}: \quad
& \underset{t_1, \cdots, t_l}{\text{maximize}}
& & \eta = \frac{\hbar_{l+1}} {\sum_{k = 1}^{l} {\prod_{i=1}^{k} (1 - \epsilon_i^{t_i}) t_k}} \\
&&& = \frac{\sum_{r = 0}^{M} {r h_{l+1,r}}} {\sum_{k = 1}^{l} {\prod_{i=1}^{k} (1 - \epsilon_i^{t_i}) t_k}}\\
& \text{subject to}
& & \text{Equation} \; (\ref{eq:rank.dist}), \\
&&& t_k \in \mathbb{N}, \quad t_k > 0, \quad k = 1, \ldots, l. \nonumber
\end{align*}

Problem $\mathbf{P1}$ is a \textit{mixed integer nonlinear programming} (MINLP) problem, which is usually NP-hard \cite{Belotti09}. Though there exist some tools (e.g., NOMAD \cite{NOMAD}, and Genetic Algorithm (GA) \cite{Gantovnik05} in MATLAB) to solve such problems, we find they cannot guarantee a good solution in reasonable time for a moderate number of intermediate nodes, e.g., $l > 5$. In particular, the solvers based on the branch-and-bound technique, such as KNITRO \cite{KNITRO}, are not suitable for problem $\mathbf{P1}$. It is due to the fact that Eq. (\ref{eq:rank.dist}) must operate on natural numbers. Therefore, we need to propose some practical methods to address our problem.
\begin{remark}
The authors in \cite{Tang16} and \cite{Yin16} illustrated that $\hbar_{l+1}$ can be further improved by making different recoding decisions, i.e., the number of coded packets of a batch, for different batches with respect to their ranks. However, these approaches inevitably bring extra computations due to the rank detection. In this paper, we are interested in achieving the maximum $\eta$ by optimizing $t_k, k = 1, \ldots, l$ without considering the rank of each batch.
\end{remark}

\section{Main Results}
In this section, we aim to efficiently solve the MINLP problem described above. We explore some inherent properties of the inner code described in Eq. (\ref{eq:inner_exp}). By combining the properties with the incomplete beta function, we establish an NLP problem, called $\mathbf{PU}$, that provides a valid upper bound of the optimal value of $\mathbf{P1}$. The upper bound is used as a direct measurement for our practical approaches. In the last part of this section, we present the centralized and decentralized optimization strategies that can produce the near-optimal solutions of $\mathbf{P1}$ in real-time.

\subsection{Properties}
Let us present in the following analysis to compute the rank distribution in a matrix fashion. For $ 1 \le k \le l, 0 \le n \le t_k$, we let
\begin{equation*}
f(k,n) = \binom{t_{k}}{n} (1 - \epsilon_k)^n \, \epsilon_k^{t_{k} - n}.
\end{equation*}

\noindent After substituting $f(k,n)$ into (\ref{eq:rank.dist}), we get
\begin{align}
\label{eq:rank.dist.inner.prod}
h_{k+1,j}
= & \sum\limits_{m=j}^{M} {\sum\limits_{n=j}^{t_k} {h_{k,m} f(k,n) \, \zeta_{j}^{m,n}}} \nonumber \\
= & h_{k,j} \sum\limits_{n=j}^{t_k} f(k,n) \zeta_{j}^{j,n} + h_{k,j+1} {\sum\limits_{n=j}^{t_k} f(k,n) \zeta_{j}^{j+1,n}} \\
& + \cdots + h_{k,M} \sum\limits_{n=j}^{t_k} f(k,n) \zeta_{j}^{M,n} \nonumber \\
= & \textbf{h}_{k}^{\text{T}} \textbf{p}_{j}^{k}, \quad 1 \le k \le l, \quad 0 \le j \le M,
\end{align}

\noindent where recall that $\textbf{h}_{k}^{\text{T}}= [h_{k,0}, h_{k,1}, \ldots, h_{k,M}]$ is the empirical rank distribution of the transfer matrices as stated above Eq.~(\ref{eq:batch_transmission_efficiency}) and, then, for $j = 1, 2, \cdots, M$,
\begin{align*}
\textbf{p}_{j}^{k} =  \left[0, \, \ldots, \, 0, \, \sum_{n=j}^{t_k} f(k,n) \zeta_{j}^{j,n}, \, \ldots, \, \sum_{n=j}^{t_k} f(k,n) \zeta_{j}^{M,n}\right],
\end{align*}
otherwise, for $j=0$,
$$ \textbf{p}_{0}^{k} = \left[1, \, \sum_{n=0}^{t_k} f(k,n) \zeta_{0}^{1,n}, \, \ldots, \, \sum_{n=0}^{t_k} f(k,n) \zeta_{0}^{M,n}\right].$$

The $(i+1)$-th component of $ \textbf{p}_{j}^{k} $ represents the probability of receiving a transfer matrix with rank $j$ given the transmitted matrix has rank $i$. We, then, model the channel as an one-step Markov process with an $(M + 1) \times (M + 1)$ \textit{transition matrix},
\begin{align}
\label{eq:transition.matrix}
\mathbf{P}_k
& \triangleq \left[ \textbf{p}_{0}^{k} \quad \textbf{p}_{1}^{k} \quad \cdots \quad \textbf{p}_{M}^{k} \right] \nonumber \\
& = \left[ \begin{array}{cccc}
1 & 0 & \cdots & 0\\
\sum\limits_{n=0}^{t_k} f(k,n) \zeta_{0}^{1,n} & \sum\limits_{n=1}^{t_k} f(k,n) \zeta_{1}^{1,n} & \cdots & 0\\
\vdots & \vdots & \ddots & \vdots\\
\sum\limits_{n=0}^{t_k} f(k,n) \zeta_{0}^{M,n} & \sum\limits_{n=1}^{t_k} f(k,n) \zeta_{1}^{M,n} & \cdots & \sum\limits_{n=M}^{t_k} f(k,n) \zeta_{M}^{M,n}
\end{array} \right], \quad 1 \le k \le l.
\end{align}

By combining (\ref{eq:rank.dist.inner.prod}) and (\ref{eq:transition.matrix}), we establish the following matrix formulas to estimate the rank distribution and the average rank at node $v_{k+1}$, respectively, for $k = 1,2, \ldots, l$ as follows:
\begin{align}
\label{eq:rank.dist.matrix}
& \textbf{h}_{k+1}^\text{T} = \textbf{h}_{k}^\text{T} \, \mathbf{P}_k = \textbf{h}_{1}^\text{T} \, \prod_{i=1}^{k} \mathbf{P}_i,
\end{align}
\noindent and
\begin{align}
\label{eq:habr.matrix}
& \hbar_{k+1} = \sum_{r=1}^{M} r h_{k+1,r} = \textbf{h}_{k+1}^\text{T} \textbf{e} = \textbf{h}_{1}^\text{T} \, \prod_{i=1}^{k} \mathbf{P}_i \textbf{e}.
\end{align}

\noindent where let $\textbf{e} = [0, 1, \ldots, M]$.

Next, we investigate the properties of the transition matrices. The following lemma indicates that the transition matrices $\mathbf{P}_k, k = 1, 2, \ldots, l$, have the same eigenvectors, while the proof is deferred in Appendix \ref{app-B}.

\begin{lemma}
\label{le:1}
\textit{
Each matrix $\mathbf{P}_k, 1 \le k \le l$, can be eigendecomposed into
}
\end{lemma}
\begin{equation}
\label{eq:eigendecompose}
\mathbf{P}_k = \mathbf{Q} \mathbf{\Lambda}_k \mathbf{Q}^{-1},
\end{equation}

\noindent where $\mathbf{\Lambda}_k$ is a diagonal matrix with eigenvalues
\begin{equation}
\label{eq:eigenvalue}
\lambda_{k,j} =
\begin{cases}
1 & j = 1, \\
\sum_{n=j - 1}^{t_k} f(k,n) \zeta_{j-1}^{j-1,n} & j = 2, 3, \ldots, M+1,
\end{cases}
\end{equation}

\noindent and $\mathbf{Q}=[q_{i,j}]_{1\leq i,j\leq M+1}$ is a lower-triangular matrix with entries
\begin{equation}
\label{eq:eigenspace}
q_{i,j} =
\begin{cases}
0 & i < j, \\
1 & j = 1, \\
\zeta_{j-1}^{i-1} & \text{otherwise}
\end{cases}
\end{equation}

With the above discussion , it is convenient to reformulate Theorem \ref{th:rank_dist} as follows.
\begin{theorem}
\label{th:rank_dist_matrix}
\textit{
The rank distribution $\mathbf{h}_{k+1}$ and the expected rank $\hbar_{k+1}$  can be derived by
}
\end{theorem}
\begin{align}
\label{eq:rank.dist.eigen}
& \mathbf{h}_{k+1}^{\text{T}} = \mathbf{h}_1^{\text{T}} \mathbf{Q} \prod_{i=1}^{k} \mathbf{\Lambda}_i \mathbf{Q}^{-1}, \\
\label{eq:hbar.eigen}
& \hbar_{k+1} = \mathbf{h}_{1}^\text{T} \, \mathbf{Q} \prod_{i=1}^{k} \mathbf{\Lambda}_i \mathbf{Q}^{-1} \textbf{e}, \quad k = 1, 2, \ldots, l,
\end{align}

\noindent \textit{where $\mathbf{\Lambda}_k$ and $\mathbf{Q}$ equal to (\ref{eq:eigenvalue}) and (\ref{eq:eigenspace}), respectively, and recall that $\mathbf{e} = [0, 1, \ldots, M]$ and $ \mathbf{h}_{1} = [0, 0, \ldots, 0, 1]$.}

Now, let $\mathbf{h_1^\text{T} Q} = [\alpha_0, \alpha_1, \ldots, \alpha_{M}]$ and $\mathbf{Q^{-1} e} = [0, \beta_1, \beta_2, \ldots, \beta_{M}]^{\text{T}}$. We transform $\mathbf{P1}$ into
\begin{align}
\mathbf{P1'}: \quad
& \underset{t_1, \cdots, t_l}{\text{maximize}}
& & \eta = \frac{\sum_{r=1}^{M} \alpha_r \beta_r \prod_{k=1}^{l} \sum_{n=r}^{t_k} f(k,n) \zeta_r^{r,n}} {\sum_{k = 1}^{l} {\prod_{i=1}^{k} (1 - \epsilon_i^{t_i}) t_k}}, \nonumber\\
& \text{subject to}
& & t_k \in \mathbb{N}, \quad t_k > 0, \quad k = 1, \ldots, l. \nonumber
\end{align}

Notice that problem $\mathbf{P1'}$ is still an MINLP.

\subsection{Upper Bound}

Here, we relax the integer restrictions of $\mathbf{P1'}$ by using the regularized incomplete beta function \cite{Zelen72}. We first give the following lemma.

\begin{lemma}
\label{le:2}
\textit{
Let $i$, $m$ and $n$ be three nonnegative integers with $i\leq \min\{m,n\}$. Then
}
\end{lemma}
\begin{align}
\label{eq:le4.1}
& \lim_{q \rightarrow \infty} \zeta_i^n = 1,
\end{align}
\noindent and
\begin{align}
\label{eq:le4.2}
& \lim_{q \rightarrow \infty} \zeta_i^{m,n} =
\begin{cases}
1 & i = \min \{ m,n \}, \\
0 & \text{otherwise}.
\end{cases}
\end{align}

\begin{IEEEproof}
Equation~(\ref{eq:le4.1}) is obvious by Eq. \eqref{zeta_r_n} in Theorem~\ref{th:rank_dist}.

For Eq. (\ref{eq:le4.2}), if $i = \min \{ m,n \}$,
\begin{align*}
\zeta_i^{m,n} =
\begin{cases}
\zeta_m^{n}, & \text{ if } m<n , \\
\zeta_n^{m}, & \text{otherwise},
\end{cases}
\end{align*}
which implies that Eq.~(\ref{eq:le4.2}) by Eq.~(\ref{eq:le4.1}). Otherwise, by Eqs. \eqref{zeta_r_n_m} and  \eqref{zeta_r_n} in Theorem~\ref{th:rank_dist}, we have
\begin{align*}
\zeta_{i}^{m,n} & = \frac{\zeta_i^n \zeta_i^m}{\zeta_i^i q^{(m-i)(n-i)}} \\
& =\dfrac{\prod_{x=0}^{i-1}(1-q^{-n+x}) \prod_{x=0}^{i-1}(1-q^{-m+x})}{\prod_{x=0}^{i-1}(1-q^{-i+x})q^{(m-i)(n-i)}},
\end{align*}
which immediately implies that $\lim_{q \rightarrow \infty} \zeta_i^{m,n} = 0$, completing the proof.
\end{IEEEproof}

Next, let us rewrite Eq. (\ref{eq:rank.dist}) as
\begin{align}
\label{eq:reform.rank.dist}
h_{k+1,j} = & \sum_{m=j+1}^{M} h_{k,m} f(k,j) \zeta_j^{m,j} + h_{k,j} \sum_{n=j}^{t_k} f(k,n) \zeta_j^{j,n} \nonumber \\
& + \sum_{\substack{m = j + 1}}^{M} \sum_{\substack{n = j + 1}}^{t_k} h_{k,m} f(k,n) \zeta_j^{m,n}.
\end{align}

Let $\widetilde{\mathbf{h}}_{k+1} = [\widetilde{h}_{k,0}, \widetilde{h}_{k,1}, \ldots, \widetilde{h}_{k,M}] $. By assuming $q$ being sufficiently large and applying Lemma \ref{le:2} to Eq. (\ref{eq:reform.rank.dist}), we, then, use the following equation to approximately calculate the rank distribution $\mathbf{h}_{k+1}$,
\begin{align}
\label{eq:apprx.rank.dist}
\widetilde{h}_{k+1,j} =
\sum_{m=j+1}^{M} \widetilde{h}_{k,m} f(k,j) + \widetilde{h}_{k,j} \sum_{n=j}^{t_k} f(k,n).
\end{align}

In particular, $\widetilde{\mathbf{h}}_{1} = \mathbf{h}_1$\footnote{When $q \ge 2^4$, the difference between $\hbar_k$ and $\widetilde{\hbar}_{k}, k > 0 $, is less than $\mathcal{O}(10^{-2})$  }. Similar to Eqs. (\ref{eq:rank.dist.matrix}) and (\ref{eq:habr.matrix}), the distribution $\mathbf{\widetilde{h}}_{k+1}$ and the corresponding expected $\mathbf{\widetilde{\hbar}}_{k+1} \triangleq \sum_{r=1}^{M} r \widetilde{h}_{k,r}$ can be expressed, respectively, as
\begin{align}
& \widetilde{\mathbf{h}}_{k+1}^\text{T} = \widetilde{\mathbf{h}}_{k}^\text{T} \, \widetilde{\mathbf{P}}_k = \widetilde{\mathbf{h}}_{1}^\text{T} \, \prod_{i=1}^{k} \widetilde{\mathbf{P}}_i, \; k = 1, \ldots, l,
\end{align}
\noindent and
\begin{align}
\label{eq:apprx.hbar.eigen}
& \widetilde{\hbar}_{k+1} = \widetilde{h}_{k+1}^\text{T} \textbf{e} = \widetilde{h}_{1}^\text{T} \, \prod_{i=1}^{k} \widetilde{\mathbf{P}}_i \textbf{e}.
\end{align}

\noindent where, the transition matrices $\widetilde{\mathbf{P}}_k,  k = 1, 2, \ldots l,$ are given by
\begin{align}
\label{eq:approx.transition.mtrx}
\widetilde{\mathbf{P}}_k \triangleq
\left[ \begin{array}{ccccc}
1 & 0 & \cdots & 0 & 0\\
\epsilon^{t_k} & \sum\limits_{n=1}^{t_k} f(k,n) & \cdots & 0 & 0\\
\epsilon^{t_k} & f(k,1) & \cdots & 0 & 0 \\
\vdots & \vdots & \ddots & \vdots & \vdots\\
\epsilon^{t_k} & f(k,1) & \cdots & f(k,M-1) &\sum\limits_{n=M}^{t_k} f(k,n)
\end{array} \right]. \nonumber
\end{align}

Then, we have the following theorem.
\begin{theorem}
\label{th:4}
\textit{
The approximate expected rank $\mathbf{\widetilde{\hbar}}_{k+1}$ can be derived by
}
\end{theorem}
\begin{equation}
\label{eq:hbar.beta}
\widetilde{\hbar}_{k+1} = \sum_{r=1}^{M} \prod_{j=1}^{k} I_{1-\epsilon_j} (r, t_j - r +1), \quad t_j \in \mathbb{N}.
\end{equation}

\noindent where
\begin{align*}
\label{eq:RIBF}
I_{1-\epsilon_j} (r, t_j - r + 1)
& = \frac{\int_{0}^{1-\epsilon_j} x^{r-1}(1-x)^{t_j-r} dx}{\int_{0}^{1} x^{r-1}(1-x)^{t_j-r} dx}
\end{align*}

\begin{IEEEproof}
Similar to Lemma \ref{le:1}, it can be verified that
\begin{equation*}
\mathbf{\widetilde{P}}_k = \mathbf{\widetilde{Q}} \mathbf{\widetilde{\Lambda}}_k \mathbf{\widetilde{Q}}^{-1},
\end{equation*}

\noindent where $\mathbf{\widetilde{\Lambda}}_k$ is a diagonal matrix with eigenvalues
\begin{equation}
\widetilde{\lambda}_{k,j} =
\begin{cases}
1 & j = 1, \\
\sum_{n=j - 1}^{t_k} f(k,n) & j = 2, 3, \ldots, M+1, \nonumber
\end{cases}
\end{equation}

\noindent and $\mathbf{\widetilde{Q}}$ is a sum matrix whose inverse $\mathbf{\widetilde{Q}}^{-1}$ is a difference matrix,
\begin{align*}
\widetilde{\mathbf{Q}} =
\left[ \begin{array}{cccccc}
1 & 0 & 0 & \cdots & 0 & 0\\
1 & 1 & 0 & \cdots & 0 & 0\\
\vdots & \vdots & \ddots & \vdots & \vdots\\
1 & 1 & 1 & \cdots & 1 & 0\\
1 & 1 & 1 & \cdots & 1 & 1
\end{array} \right]
\end{align*}
\noindent and
\begin{align*}
\widetilde{\mathbf{Q}}^{-1} =
\left[ \begin{array}{cccccc}
1 & 0 & 0 & \cdots & 0 & 0\\
-1& 1 & 0 & \cdots & 0 & 0\\
0 &-1 & 1 & \cdots & 0 & 0\\
\vdots & \vdots & \vdots & \ddots & \vdots & \vdots\\
0 & 0 & 0 & \cdots &-1 & 1
\end{array} \right].
\end{align*}

In particular, note that
\begin{align}
& \mathbf{h}_1 \widetilde{\mathbf{Q}}  = [1,1,\ldots,1], \nonumber \\
& \widetilde{\mathbf{Q}}^{-1} \mathbf{e} = [0,1,1,\ldots,1]^{\text{T}}. \nonumber
\end{align}

Then, by Eq.~(\ref{eq:apprx.hbar.eigen}) we obtain that for $k = 1, 2, \ldots, l$,
\begin{align*}
\widetilde{\hbar}_{k+1}
& = \widetilde{\mathbf{h}}_1^{\text{T}} \mathbf{\widetilde{Q}} \prod_{i=1}^{k} \mathbf{\widetilde{\Lambda}}_i \mathbf{\widetilde{Q}}^{-1} \mathbf{e} \nonumber \\
& = \sum_{r=1}^{M} \prod_{j=1}^{k} \sum_{n=r}^{t_j} f(j,n).
\end{align*}

Furthermore, by using the following property of the regularized incomplete beta function, the proof is completed.
\begin{align*}
I_{1-\epsilon_j} (r, t_j - r + 1) & = \sum_{n=r}^{t_j} \binom{t_j}{n} (1 - \epsilon_j)^n \, \epsilon_j^{t_j - n} \\
& = \sum_{n=r}^{t_j} f(j,n), \quad t_j,r \in \mathbb{N}.
\end{align*}

\end{IEEEproof}

Now, with Eq. (\ref{eq:hbar.beta}), we construct the following formulation,
\begin{align}
\label{eq:appx.batch.trans.efficiency}
\frac{\widetilde{\hbar}_{l+1}} {\sum_{k = 1}^{l} {\prod_{i=1}^{k} (1 - \epsilon_k^{t_i}) t_k}} = \frac{\sum_{r=1}^{M} \prod_{k=1}^{l} I_{1-\epsilon_i} (r, t_k - r +1)} {\sum_{k = 1}^{l} {\prod_{i=1}^{k} (1 - \epsilon_i^{t_i}) t_k}}.
\end{align}

It is worth to notice that the parameter $t_j$ in $I_{1-\epsilon_k} (r, t_k - r +1)$ does not need to be an integer. Hence, we establish a new optimization problem as follows,
\begin{align*}
\mathbf{PU}: \quad
& \underset{t_1, \cdots, t_l}{\text{maximize}}
& & \frac{\sum_{r=1}^{M} \prod_{k=1}^{l} I_{1-\epsilon_k} (r, t_k - r +1)} {\sum_{k = 1}^{l} {\prod_{i=1}^{k} (1 - \epsilon_i^{t_i}) t_k}} \\
& \text{subject to}
& & t_k > 0, \quad k = 1, \ldots, l. \nonumber
\end{align*}

For this nonlinear programming problem, we have the following theorem whose proof is deferred to Appendix \ref{app-C}.

\begin{theorem}
\label{th:5}
\textit{
Problem $\mathbf{PU}$ yields an upper bound of the optimal value of problem $\mathbf{P1}$ (or $\mathbf{P1'}$).
}
\end{theorem}

However, the solutions of Problem $\mathbf{PU}$ are not feasible for $\mathbf{P1}$ since their values in general are not integers. Therefore, we still need to develop some practical schemes to solve our problem.

\subsection{Real-Time Implementations}

First, we design a centralized approach by constructing the following optimization problem.
\begin{align}
\mathbf{P2}: \quad
& \underset{t_1, \cdots, t_l}{\text{maximize}}
& & \frac{\sum_{r=1}^{M} \alpha_r \beta_r \prod_{i=1}^{l} \sum_{n=r}^{t_i} f(i,n) \zeta_r^{r,n}} {\sum_{k = 1}^{l} {\prod_{i=1}^{k} (1 - \epsilon_i^{t_i}) t_k}}, \nonumber\\
& \text{subject to}
& & t_k^u \le t_k \le t_k^d, \quad t_k \in \mathbb{N}, \; k = 1, 2, \ldots, l. \nonumber
\end{align}

\noindent where $\mathbf{u} = [t_1^u, \ldots, t_l^u]$ and $\mathbf{d} = [t_1^d, \ldots, t_l^d]$ are computed by Algorithm \ref{alg:1}. The problem $\mathbf{P1'}$ and $\mathbf{P2}$ have the same objective function. Nevertheless, the feasible region of $\mathbf{P2}$ is finite and, more specifically, contains $2^{l}$ elements at most. Various MINLP solvers, such as NAMOD and GA, can solve $\mathbf{P2}$ in reasonable time. Thus, the source can use the network information gathered in the initial phase to compute the solutions of $\mathbf{P2}$ in real-time.

\begin{algorithm}[!h]
\caption{Finding boundary for $t_k$}
\label{alg:1}
\begin{algorithmic}[1]
\REQUIRE ~
$M$, $l$, $\epsilon_1, \epsilon_2, \ldots,\epsilon_l$;
\ENSURE ~
$\mathbf{u}$ and $\mathbf{d}$;
\STATE $\mathbf{\widetilde{t}} \gets \arg \min \widetilde{\eta}$;
\STATE $\mathbf{u} \gets \lceil\mathbf{\widetilde{t}}\rceil$;
\STATE $\mathbf{d} \gets \lfloor\mathbf{\widetilde{t}}\rfloor$;
\RETURN $\mathbf{u}$, $\mathbf{d}$;
\end{algorithmic}
\end{algorithm}

\begin{figure}
\centering
\begin{tikzpicture}[auto, outer sep=3pt, node distance=2.5cm,>=latex']
\node [end node, label=above:{18},label=below:{18}] (S) {S};
\node [relay node, right of = S, label=above:{18}, label=below:{16}] (R) {R};
\node [end node, right of = R, label=above:{$\eta = 2.68 $}, label=below:{$\eta' = 2.50$}] (D) {D};
\draw [->,thick] (S) --  node[above] {0.2} node[below] {0.2}(R) ;
\draw [->,thick] (R) --  node[above] {0.2} node[below] {0.1}(D);
\end{tikzpicture}
\caption{The 2-hop network, $M = 16$. The optimal solutions of $\mathbf{P2}$ for packet loss rates (0.2,0.2) and (0.2,0.1) are (18,18) and (18,16), respectively.}
\label{fig:3}
\end{figure}
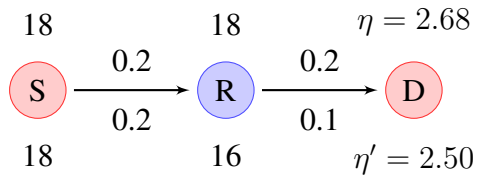

For the centralized approach, one question should be asked: how to send the solutions to the intermediate nodes. One possible way is to put them into the packet header or construct a new control packet containing the information. Either way, it inevitably takes extra overhead. Also, this global estimate may degrade the network performance in the presence of the time-varying network. For example, as shown in Fig. \ref{fig:3}, suppose PLRs change from (0.2, 0.2) to (0.2, 0.1) in the transmission phase. Using the centralized method, intermediate node $R$ is unable to timely adapt to such a change, resulting in $(18 - 16) * 100\% / 16 = 12.5\%$ more packets transmitted.

To overcome the above problems, we propose a decentralized method, such that each node makes their own inner coding decisions locally and independently. First, we have to face the question that how to solve problem $\mathbf{P1}$ (or $\mathbf{P1'}$) locally, i.e., how nodes choose the parameters $M, l, \epsilon_1, \epsilon_2, \ldots,\epsilon_l$, based on their own knowledge of the network. Obviously, the problem is trivial when every node can obtain the global information. However, in real system, nodes may only be able to gather network information within a certain range. In particular, we assume that each node can only acquire 1-hop packet loss rates from its neighbors.\footnote{Periodic Hello messaging can be used to perform this task.} Further, let us suppose that both the number of hops $l$, and the batch size $M$ are contained in packet header, which do not cost too much. Then, the remaining question is how to set packet loss rates $\epsilon_1, \ldots, \epsilon_l$. Our solution is inspired by the following facts.

We notice that the probabilities $p_i, i = 1, 2, \ldots, l$, are very close to 1, when the batch size $M \ge 8$. Therefore, we set $p_i = 1, i = 1, 2, \ldots, l$, and, then, construct the resulting optimization problem $\mathbf{PA}$.
\begin{align*}
\mathbf{PA}: \quad
& \underset{t_1, \cdots, t_l}{\text{maximize}}
& & \frac{\sum_{r=1}^{M} \alpha_r \beta_r \prod_{i=1}^{l} \sum_{n=r}^{t_i} f(i,n) \zeta_r^{r,n}} {\sum_{k = 1}^{l} {t_k}} \\
& \text{subject to}
& & t_k \in \mathbb{Z}, \quad t_k > 0, \quad k = 1, 2, \ldots, l. \nonumber
\end{align*}

\begin{remark}
The solutions of $\mathbf{PA}$ well match these of $\mathbf{P1}$, even when $q = 2$ and $M = 8$. This is because, the optimal solutions for both $\mathbf{PA}$ and $\mathbf{P1}$ are all larger than $M$ and increase with both path length and PLR. As a results, the product of $(1 - \epsilon_k^{t_i})$ will be very close to 1 in such cases. For example, given $l = 100, \epsilon_1 = \epsilon_2 = \cdots = \epsilon_l = 0.35, q = 2$ and $M = 8$, then the optimal solution for both $\mathbf{PA}$ and $\mathbf{P1}$ is $t_1^{\ast} = \cdots = t_l^{\ast} = 25$, and $\prod_{i = 1}^{100} (1 - \epsilon_k^{25}) = 0.99999991$. In the meanwhile, the difference between the objective values of $\mathbf{PA}$ and $\mathbf{P1}$ is less than $\mathcal{O}(10^{-5})$.
\end{remark}

Then, the following proposition is derived from problem $\mathbf{PA}$.

\begin{proposition}
\label{Prop:1}
\textit{
Let $\{ t_i^{\ast}, i = 1,2,\ldots,l \}$, be the optimal solution for $\mathbf{PA}$. If $\epsilon_1 = \epsilon_2 = \cdots = \epsilon_l$, then $t_1^{\ast} = t_2^{\ast} = \cdots = t_l^{\ast}$.
}
\end{proposition}

\begin{IEEEproof}
Let $\epsilon = \epsilon_i$ and $t = t_i, i = 1,2,\ldots,l$. We define
\begin{align*}
\eta'
& = \frac{\sum_{r=1}^{M} \alpha_r \beta_r \left( \sum_{n=r}^{t} \binom{t}{n} (1 - \epsilon)^n \, \epsilon^{t - n} \zeta_r^{r,n} \right)^l } {lt} \\
& = \frac{\sum_{r=1}^{M} \alpha_r \beta_r g^l(r,t)} {lt}, \quad t \in \mathbb{N} , t > 0.
\end{align*}

\noindent where $$g(r,t) = \sum_{n=r}^{t} \binom{t}{n} (1 - \epsilon)^n \, \epsilon^{t - n} \zeta_r^{r,n}.$$ Let $t^{\ast} = \arg \max_t \eta'$. Then for any feasible $t$,
\begin{align}
\label{eq:th6.2}
\frac{\sum_{r=1}^{M} \alpha_r \beta_r g^l(r,t^{\ast})} {l t^{\ast}} \ge \frac{\sum_{r=1}^{M} \alpha_r \beta_r g^l(r,t)} {l t}.
\end{align}

By taking ${t_1, t_2, \ldots, t_k}$ into Eq. (\ref{eq:th6.2}) and adding them up, we write
\begin{align*}
\frac{\sum_{r=1}^{M} \alpha_r \beta_r g^l(r,t^{\ast})}{lt^{\ast}} \left( \sum_{i=1}^{l} l t_i \right)
& \ge \sum_{r=1}^{M} \alpha_r \beta_r \left( \sum_{k=1}^{l} g^l(r,t_k) \right)\\
& \overset{(a)}{\ge} \sum_{r=1}^{M} \alpha_r \beta_r l \sqrt[\leftroot{-3}\uproot{3}l]{g^l(r,t_1)g^l(r,t_2) \cdots g^l(r,t_l)} \\
& = \sum_{r=1}^{M} \alpha_r \beta_r l \prod_{k=1}^{l} g(r,t_l).
\end{align*}

\noindent where (a) follows the arithmetic-geometric mean inequality. The above equation implies
\begin{align*}
\frac{\sum_{r=1}^{M} \alpha_r \beta_r g^l(r,t^{\ast})}{lt^{\ast}} \ge \frac{\sum_{r=1}^{M} \alpha_r \beta_r \prod_{k=1}^{l} g(r,t_l)}{\sum_{k = 1}^{l} t_k}, \\
\forall t_i > 0, \; i = 1, 2,\ldots, l
\end{align*}

The proof is completed.
\end{IEEEproof}

\begin{corollary}
\label{Cor:2}
\textit{
Let $\{ t_i^{\ast}, i = 1,2,\ldots,l \}$, be the optimal solution for $\mathbf{PA}$. If $\epsilon_{i_1} = \epsilon_{i_2} = \cdots = \epsilon_{i_n}$, then $t_{i_1}^{\ast} = t_{i_2}^{\ast} = \cdots = t_{i_n}^{\ast}$, $1 \le i_k \le l$, $1 < k \le n \le l$.
}
\end{corollary}

\begin{IEEEproof}
We consider $n < l$ and define two index sets $\Omega = \{ 1, 2, \ldots, l \}$ and $\bar{\Omega} = \{ i_1, i_2, \ldots, i_n \}$. Then, the objective function of $\mathbf{PA}$ can be rewritten as follows
\begin{align}
\eta' \triangleq \frac{T + \sum_{i \in \bar{\Omega}} t_i} { \sum_{r=1}^{M} \alpha_r \beta_r \gamma_r \prod_{i \in \bar{\Omega}} \sum_{n=r}^{t_i} f(i,n) \zeta_r^{r,n}}, \nonumber
\end{align}

\noindent where $T = \sum_{i \in \Omega / \bar{\Omega}} t_i^{\ast}$ and $\gamma_r = \prod_{i \in \Omega / \bar{\Omega}} \sum_{n=r}^{t_i^{\ast}} f(i,n) \zeta_r^{r,n}$ are constants. Now, we can apply the method in Proposition \ref{Prop:1} to prove this corollary.
\end{IEEEproof}

Subsequently, we construct a single variable optimization problem as follows.
\begin{align*}
\mathbf{PS}: \quad
& \underset{t}{\text{maximize}}
& & \frac{\sum_{r=1}^{M} \alpha_r \beta_r \left( \sum_{n=r}^{t} \binom{t}{n} (1 - \epsilon)^n \, \epsilon^{t - n} \zeta_r^{r,n} \right)^l} {l t}\nonumber\\
& \text{subject to}
& & t > 0, t \in \mathbb{Z}. \nonumber
\end{align*}

Nodes $v_{i}, i = 1,2,\ldots,l$, can input $\epsilon = \epsilon_i$ to $\mathbf{PS}$, and use the outputs to make their own coding decisions. From Table \ref{tab:1}, of which each element is the solution for $\mathbf{PS}$ with a pair of PLR and the number of hops, we observe that the difference between any two adjacent elements is at most 1. It suggests that we can build a 4-D look-up table for $q$, $M$, and use a fixed step size of 0.01 and 1 for $\epsilon$ and $l$, respectively. Therefore, nodes can use the look-up table to make the appropriate decisions by matching the current situation, i.e., $q$, $M$, $l$ and $\epsilon$, to the most similar of the entries already in the table. The time complexity of this table look-up algorithm is only $\mathcal{O}(1)$ at each node. In addition, Table \ref{tab:1} can be further compressed by eliminating the repeated elements.

\begin{table*}
\centering
\caption{The look-up table, $q = 2^8$, $M = 16$}
\label{tab:1}
\begin{tabular}{{l}*{19}{c}}    %
\toprule
\multicolumn{1}{l}{} & \multicolumn{19}{c}{\bf The number of hops} \\
\cmidrule{2-20}
\multicolumn{1}{l}{\bf PLR} & 2 & 3 & 4 & 5 & 6 & 7 & 8 & 9 & 10 & 11 & 12 & 13 & 14 & 15 & 16 & 17 & 18 & 19 & 20\\
\midrule
0.10 & 16 & 17 & 17 & 18 & 18 & 18 & 18 & 19 & 19 & 19 & 19 & 19 & 19 & 19 & 19 & 19 & 19 & 20 & 20 \\
\rowcolor{mygray}
0.11 & 17 & 17 & 18 & 18 & 18 & 19 & 19 & 19 & 19 & 19 & 19 & 19 & 20 & 20 & 20 & 20 & 20 & 20 & 20 \\
0.12 & 17 & 17 & 18 & 18 & 19 & 19 & 19 & 19 & 19 & 19 & 20 & 20 & 20 & 20 & 20 & 20 & 20 & 20 & 20 \\
\rowcolor{mygray}
0.13 & 17 & 18 & 18 & 19 & 19 & 19 & 19 & 19 & 20 & 20 & 20 & 20 & 20 & 20 & 20 & 20 & 20 & 21 & 21 \\
0.14 & 17 & 18 & 18 & 19 & 19 & 19 & 20 & 20 & 20 & 20 & 20 & 20 & 20 & 21 & 21 & 21 & 21 & 21 & 21 \\
\rowcolor{mygray}
0.15 & 17 & 18 & 19 & 19 & 19 & 20 & 20 & 20 & 20 & 20 & 21 & 21 & 21 & 21 & 21 & 21 & 21 & 21 & 21 \\
0.16 & 17 & 18 & 19 & 19 & 20 & 20 & 20 & 20 & 21 & 21 & 21 & 21 & 21 & 21 & 21 & 21 & 22 & 22 & 22 \\
\rowcolor{mygray}
0.17 & 17 & 18 & 19 & 20 & 20 & 20 & 20 & 21 & 21 & 21 & 21 & 21 & 21 & 22 & 22 & 22 & 22 & 22 & 22 \\
0.18 & 18 & 19 & 19 & 20 & 20 & 21 & 21 & 21 & 21 & 21 & 22 & 22 & 22 & 22 & 22 & 22 & 22 & 22 & 22 \\
\rowcolor{mygray}
0.19 & 18 & 19 & 20 & 20 & 21 & 21 & 21 & 21 & 22 & 22 & 22 & 22 & 22 & 22 & 22 & 22 & 23 & 23 & 23 \\
0.20 & 18 & 19 & 20 & 20 & 21 & 21 & 21 & 22 & 22 & 22 & 22 & 22 & 23 & 23 & 23 & 23 & 23 & 23 & 23 \\
\bottomrule
\end{tabular}
\end{table*}

\begin{remark}
There are many choices to set packet loss rates $\epsilon_1, \epsilon_2, \ldots, \epsilon_l$ and construct the corresponding look-up tables. For example, since node $v_i$ knows $\epsilon_{i-1}$, we can set $v_k = v_i, k = 1,2,\ldots,l, k \neq i - 1$. As another example, since node $v_i$ can computer the empirical rank distribution of the batches it received, only $\epsilon_k, k = i+1,\ldots,l$, need be set. However, these methods may lead to a very large table costing a lot of resources. As we shall see in the next section, the proposed method already achieves a near-optimal performance by comparing with the upper bound.
\end{remark}

In the above discussion, we assume that the number of hops, $l$, can be determined precisely by routing policy. However, in real systems, the routing protocol may only return an approximate value of $l$, or the intermediate nodes may change in the transmission phase. It results in lengthening or shortening the path length. Therefore, setting $l$ to be an fixed value may cause inaccuracy. Our solution of this problem is based on the following facts.

\begin{proposition}
\label{Prop:2}
\textit{
Given $\epsilon$, postulate that $t_1^{\ast}$ and $t_2^{\ast}$ are the optimal solutions for $\mathbf{PS}$ with parameters $l_1$ and $l_2$, respectively. If $l_1 < l_2$, then $t_1^{\ast} \le t_2^{\ast}$.
}
\end{proposition}
\begin{IEEEproof}
The proof is given in Appendix \ref{app-D}.
\end{IEEEproof}

\begin{corollary}
\label{col:2}
\textit{
Given $\epsilon_1, \epsilon_2,\ldots, \epsilon_l$, $t_1, t_2,\ldots, t_l$ and $\hat{t}_1, \hat{t}_2,\ldots, \hat{t}_l$. If $t_i \le \hat{t}_i$, $\forall i = 1,2,\ldots,l$, then $$\Pr \{ {\rm{rk}} (\mathbf{H}_i) \ge r \} \ge \Pr \{ {\rm{rk}} (\mathbf{\hat{H}}_i) \ge r \},$$ where the transfer matrices $\mathbf{H}_i$ and $\mathbf{\hat{H}}_i$ are related to $t_1, t_2,\ldots, t_i$ and $\hat{t}_1, \hat{t}_2,\ldots, \hat{t}_i$, respectively.
}
\end{corollary}

The above results suggest that we can set a range of the number of hops for a path, and choose the largest value among them as the parameter $l$ inserted into packet headers. In this way, it can provide a ``good'' rank distribution (higher average rank) if the length of a path is within the range. In this paper, we construct the look-up tables, called \textit{Refined Lookup Table} (RLT), by only keeping the $2^{nd}, 4^{th}, 7^{th}, 11^{th}, 16^{th}$ and $20^{th}$ columns of the CLTs. Table II gives an instance of RLTs. With RLTs, the source will build the nearest (and larger) $l$ listed in RLT into packet headers according to the routing information. For example, the value $l = 7$ will be inserted into packet headers, if the number of hops reported by the routing protocol is larger than 2 but no more than 7. In the next section, the simulation results will show that the use of RLT causes only a little loss in the performance.

\begin{table}
\centering
\caption{The refined look-up table, $q = 2^8$, $M = 16$}
\label{tab:2}
\begin{tabular}{{l}*{6}{c}}    %
\toprule
\multicolumn{1}{l}{} & \multicolumn{6}{c}{\bf The number of hops} \\
\cmidrule{2-7}
\multicolumn{1}{l}{\bf PLR} & 2 & 4 & 7 & 11 & 16& 20\\
\midrule
0.10 & 16 & 17 & 18 & 19 & 19 & 20 \\
\rowcolor{mygray}
0.11 & 17 & 18 & 19 & 19 & 20 & 20 \\
0.12 & 17 & 18 & 19 & 19 & 20 & 20 \\
\rowcolor{mygray}
0.13 & 17 & 18 & 19 & 20 & 20 & 21 \\
0.14 & 17 & 18 & 19 & 20 & 21 & 21 \\
\rowcolor{mygray}
0.15 & 17 & 19 & 20 & 20 & 21 & 21 \\
0.16 & 17 & 19 & 20 & 21 & 21 & 22 \\
\rowcolor{mygray}
0.17 & 17 & 19 & 20 & 21 & 22 & 22 \\
0.18 & 18 & 19 & 21 & 21 & 22 & 22 \\
\rowcolor{mygray}
0.19 & 18 & 20 & 21 & 22 & 22 & 23 \\
0.20 & 18 & 20 & 21 & 22 & 23 & 23 \\
\bottomrule
\end{tabular}
\end{table}

\section{Numerical Results and Discussion}

In this section, we present numerical experiments to evaluate the performance of the proposed algorithms, and demonstrate the improvement compared to the original BATS codes \cite{YangTIT14}.

The unicast wireless networks described in Section III is considered. We assume that the links are heterogeneous. More specifically, each node may experience different packet loss rate (PLR). In the experiments, network size $l$ ranges from 2 to 20, and  batch size $M$ is set to 12, 16, 20 and 24, respectively. For a combination of $l$ and $M$ (e.g $l = 2, M = 12$), we run 10000 trials, in each of which packet loss rates are randomly chosen from a uniform distribution over the interval [0.05,0.35]. The loss rates remain constant over each trial. Figures are, then, derived from average analytical results. The simulations are run in MATLAB, and all optimization problems are solved by means of the optimization toolbox.

In particular, we separately test two classes of the look-up tables for the decentralized approach. One uses a fixed step size of 1 for the number of hops, called complete look-up tables, such as Table \ref{tab:1}. Another class of tables, called refined look-up table, contains only the $2^{nd}, 4^{th}, 7^{th}, 11^{th}, 16^{th}$ and $20^{th}$ columns of the CLTs.

\subsection{Transmission Efficiency}

\begin{figure}[!t]
\center
\subfloat[Transmission efficiency]{\includegraphics[width=3.1in]{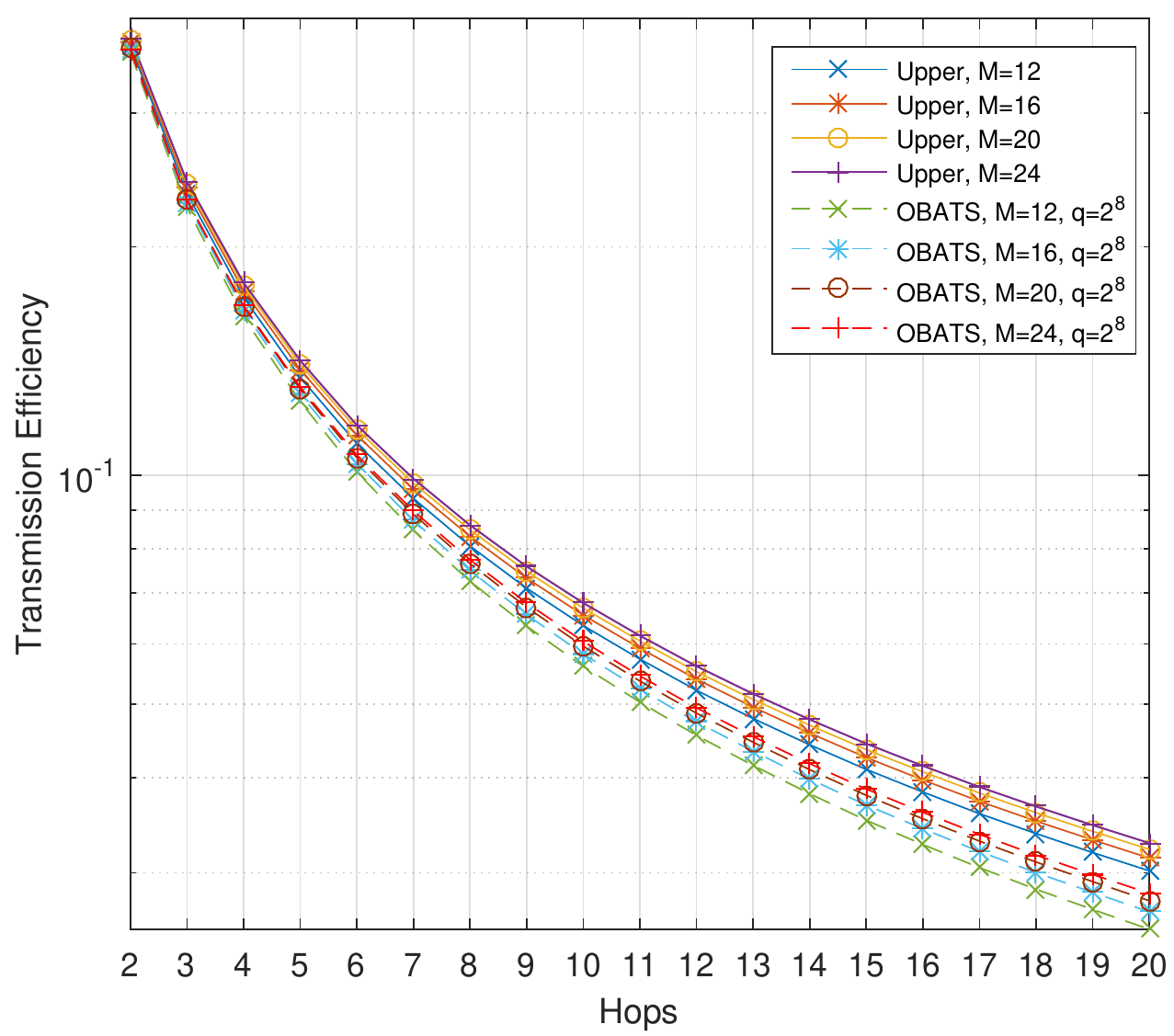}
\label{trans efficiency_upper_vs_obats_a}}
\hfil
\subfloat[Reduction compared to the original BATS]{\includegraphics[width=3.3in]{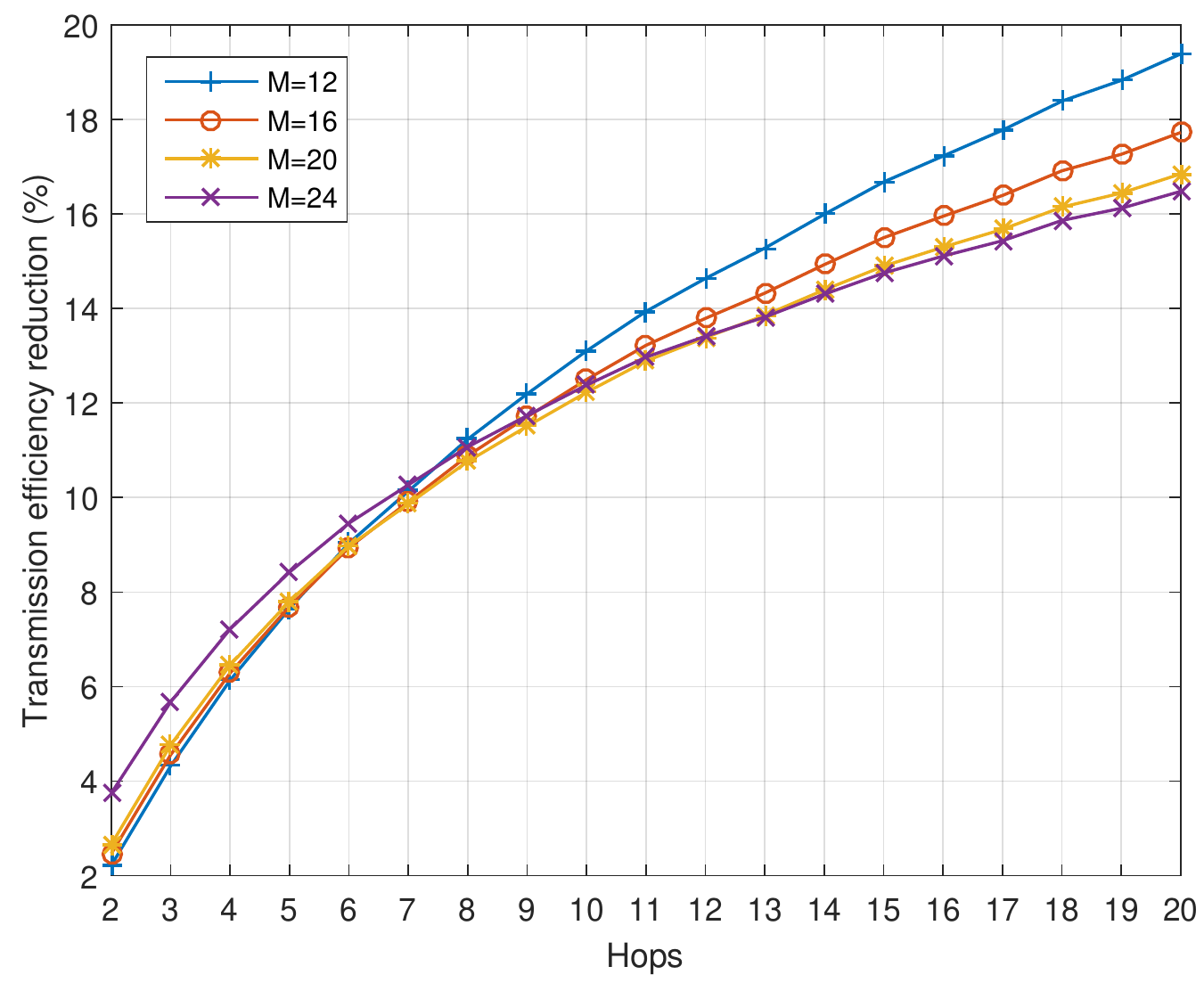}
\label{trans efficiency_upper_vs_obats_b}}
\hfil
\caption{Transmission efficiency vs the number of hops under different batch size.}
\label{trans efficiency_upper_vs_obats}
\end{figure}

In the first set of experiments, we would like to check the performance of the original BATS (denoted by OBATS) codes \cite{YangTIT14} and the upper bound (denoted by Upper) in terms of transmission efficiency. We set the field size $q = 2^8$. Notice that the performance of the original BATS codes can be barely improved by increasing the field size to $q \ge 2^8$. The batch transmission efficiency is depicted in Fig. \ref{trans efficiency_upper_vs_obats_a}, and more specifically, Fig. \ref{trans efficiency_upper_vs_obats_b} illustrates the percentage of improvement of transmission efficiency compared to OBATS. \textit{The transmission efficiency reduction = [Upper - OBATS] $\times$ 100 / OBATS}.

\begin{figure*}[!t]
\center
\subfloat[$M = 12$]{\includegraphics[width=3in]{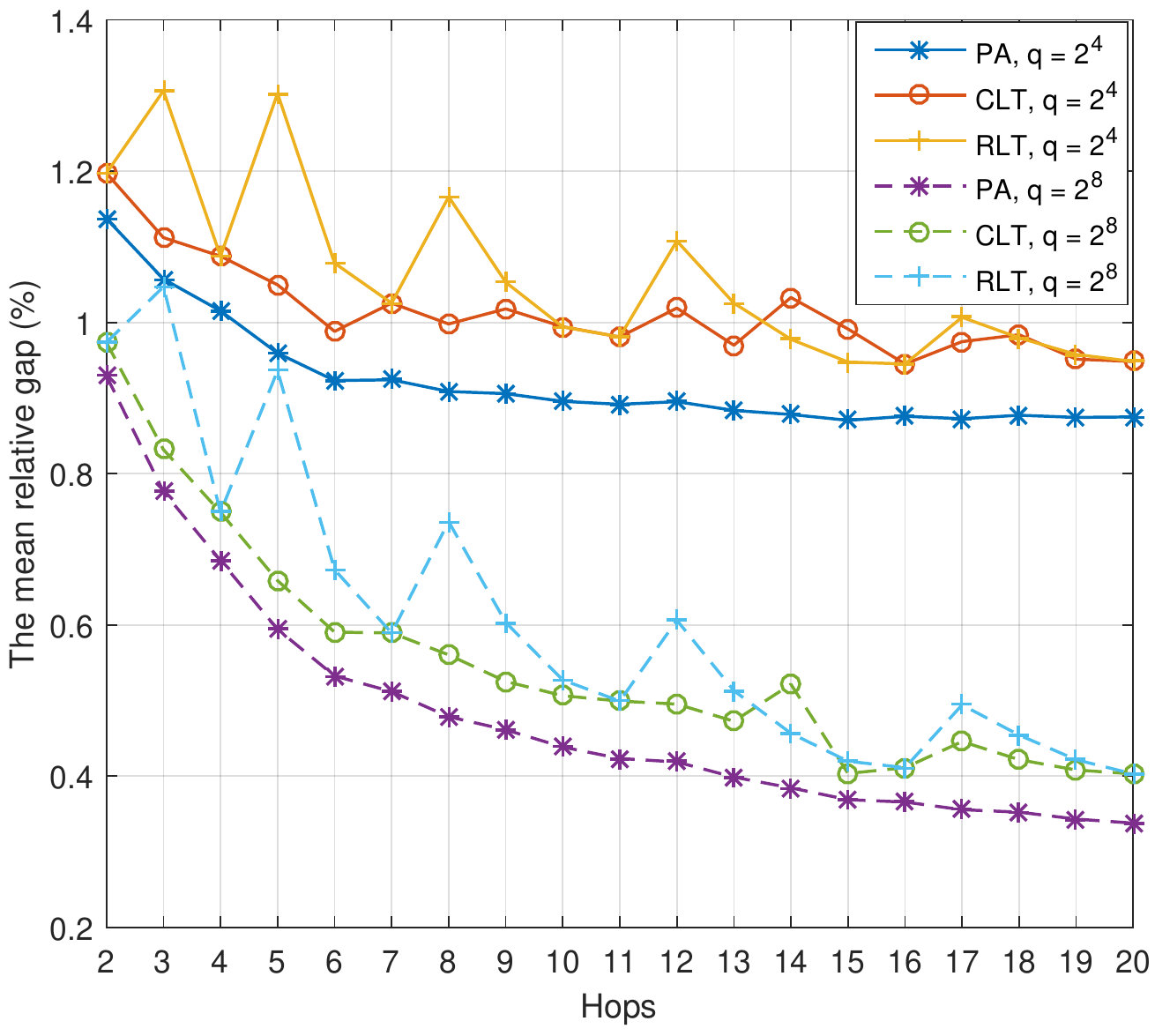}
\label{trans efficiency_gap_upper_vs_realtime_M12}}
\hfil
\subfloat[$M = 16$]{\includegraphics[width=3in]{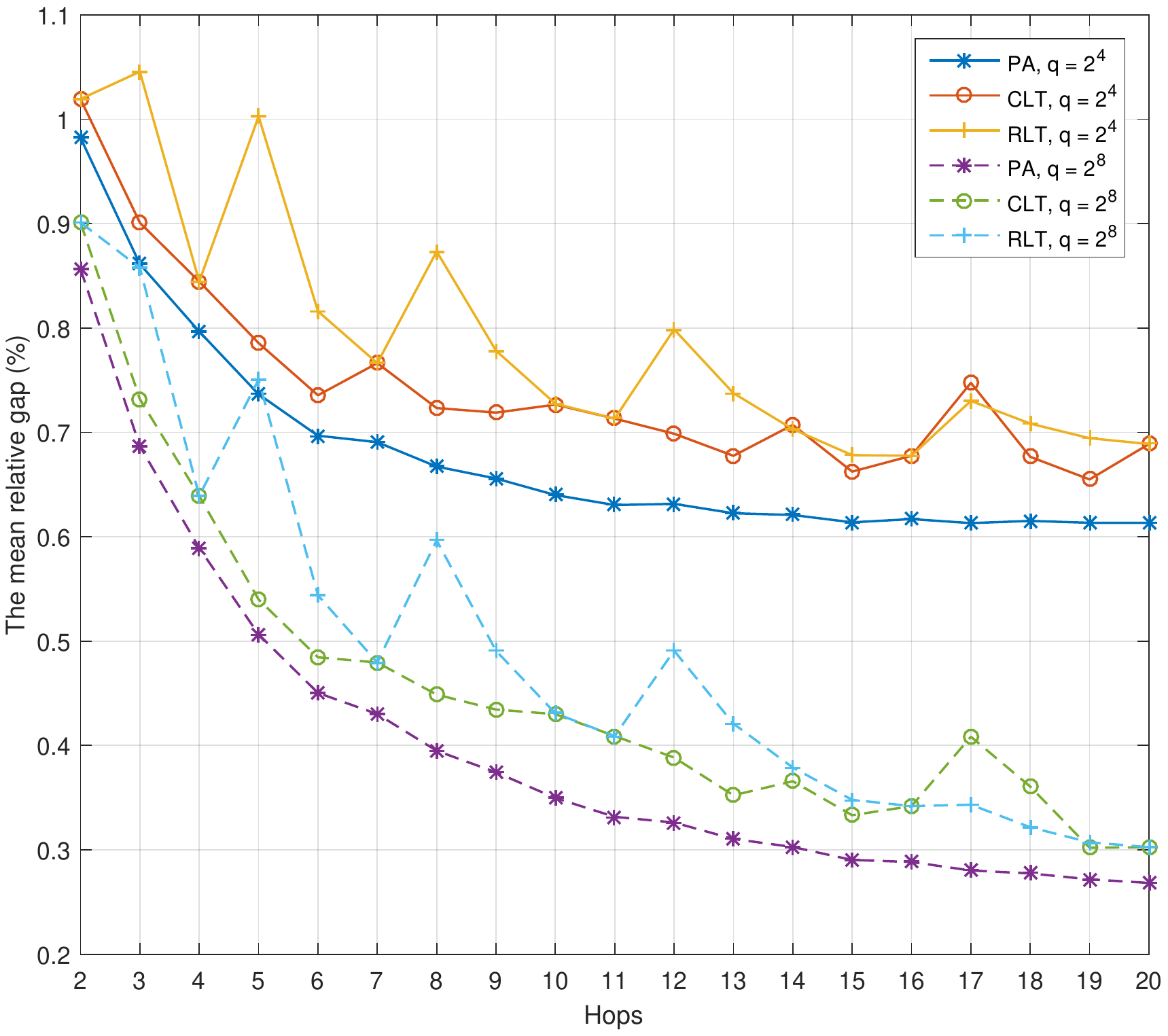}
\label{trans efficiency_gap_upper_vs_realtime_M16}}
\hfil
\subfloat[$M = 20$]{\includegraphics[width=3in]{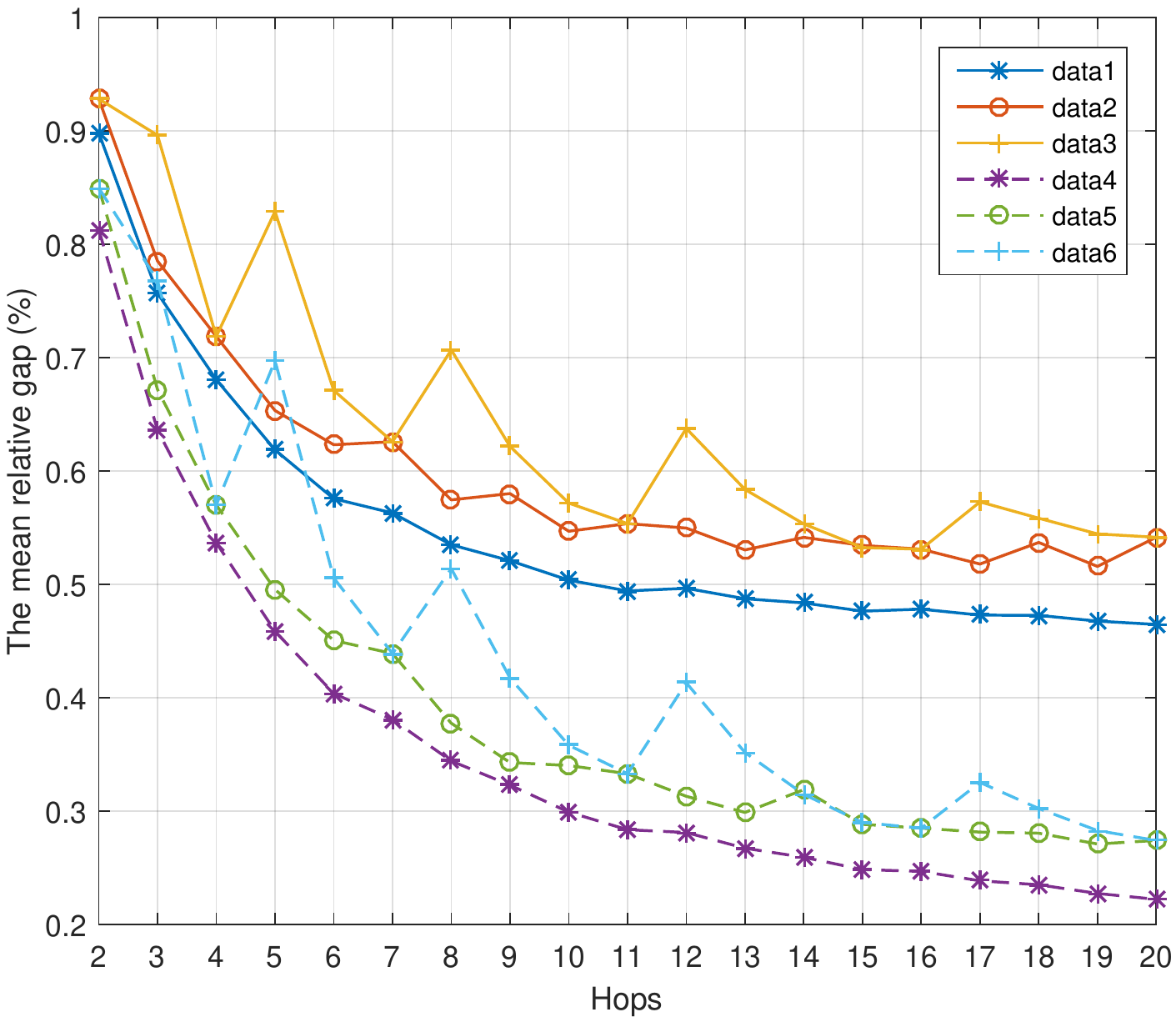}
\label{trans efficiency_gap_upper_vs_realtime_M20}}
\hfil
\subfloat[$M = 24$]{\includegraphics[width=3.1in]{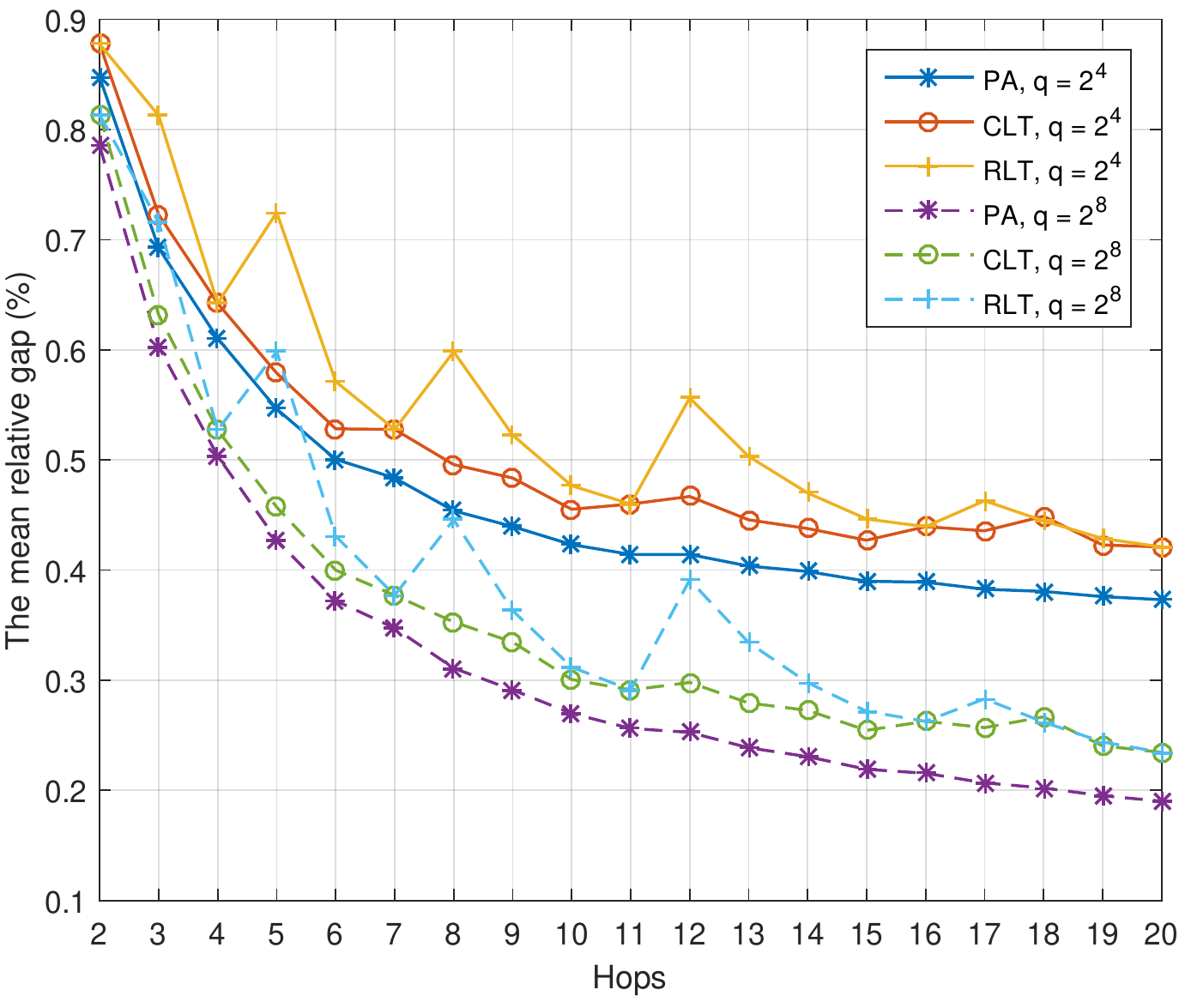}
\label{trans efficiency_gap_upper_vs_realtime_M24}}
\hfil
\caption{The mean relative gap vs the number of hops under different batch size. The lines marked by P3, CLT and RLT represent the P3-based, CLT-based, RLT-based algorithms, respectively. }
\label{trans efficiency_gap_upper_vs_realtime}
\end{figure*}

Next, we use the upper bound as the performance benchmark to test our proposed algorithms. In the simulation, we set the field size $q = 2^4$ and $q = 2^8$, respectively. We use PA, CLT and RLT to denote the centralized, the CLP-based and the RLT-based real-time approach, respectively. The \textit{mean relative gap} between the real-time approach and the upper bound is defined by \textit{[Upper - PA (CLT or RLT)] $\times$ 100 / Upper}. Results are depicted in Fig. \ref{trans efficiency_gap_upper_vs_realtime}. In general, we observe that all the approximating solutions closely match the upper bound. More importantly, the look-up-table-based approximations achieve extremely competitive performance with the upper bound, but with a constant computational cost. In particular, the transmission efficiency are very similar in the two table look-up algorithms.

Combining Fig. \ref{trans efficiency_upper_vs_obats} with Fig. \ref{trans efficiency_gap_upper_vs_realtime}, the improvement of the proposed methods can clearly seen in all cases, i.e., for all kind of batch size and network size. Therefore, according to the discussion in Section III, it confirms that the total number of transmissions can be significantly reduced by adjusting the values of $t_i, i = 1,\ldots$ adapting to the current channel conditions.

Besides, Fig. \ref{trans efficiency_gap_upper_vs_realtime} also presents that the gap between the approximating results and the upper bound gets smaller with the increase of the number of hops as well as the batch size. On the other hand, we also observed that the curves corresponding to the RLT-based approach are sharper when the path length is smaller than 12, while they become smoother at longer length. These observations illustrate that the transmission efficiency is more sensitive when the number of hops is relatively small.

\subsection{The Average Rank}

\begin{figure*}[!t]
\center
\subfloat[$M = 12$]{\includegraphics[width=3in]{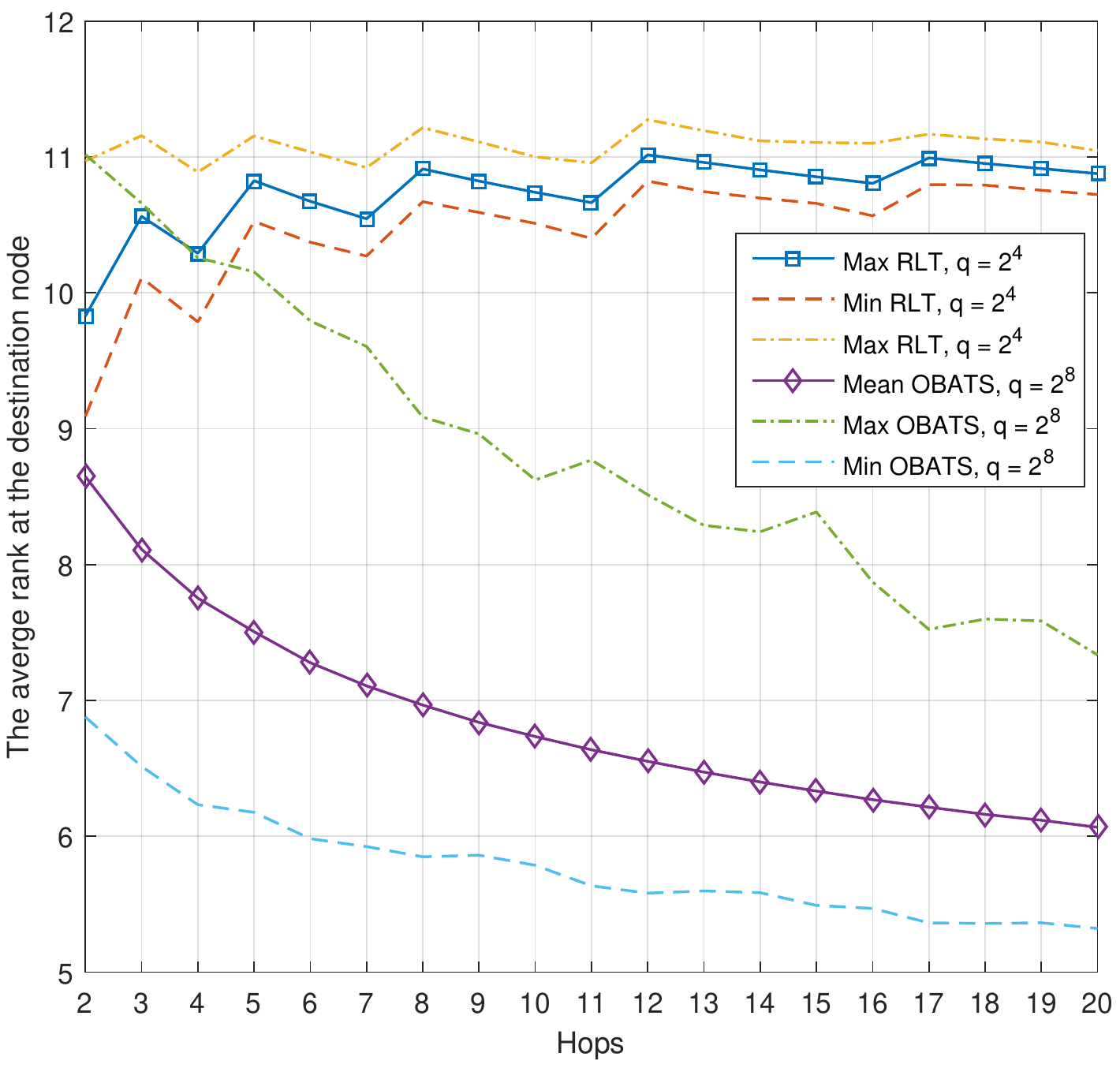}
\label{avg_rank_M12}}
\hfil
\subfloat[$M = 16$]{\includegraphics[width=3in]{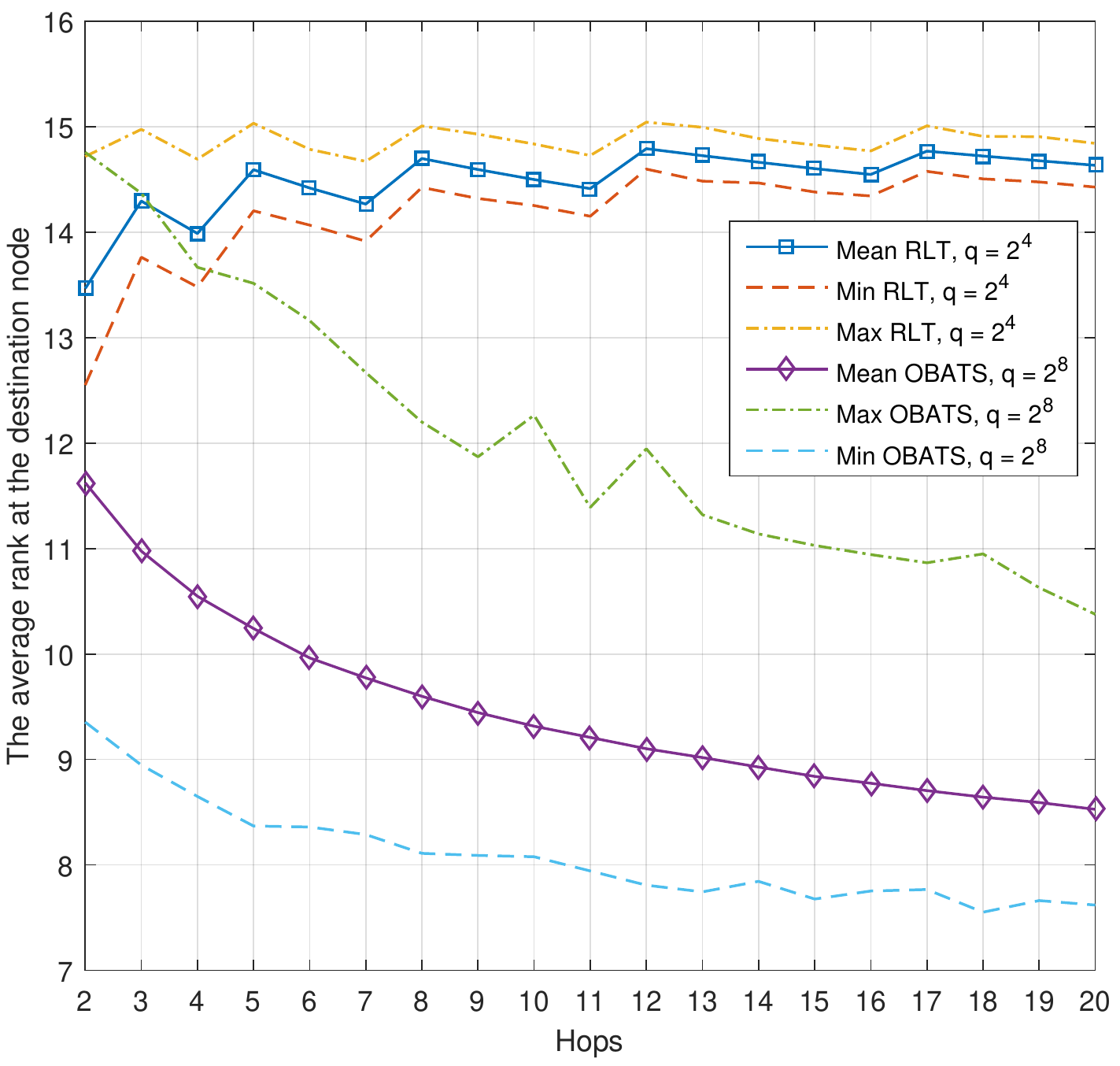}
\label{avg_rank_M16}}
\hfil
\subfloat[$M = 20$]{\includegraphics[width=3in]{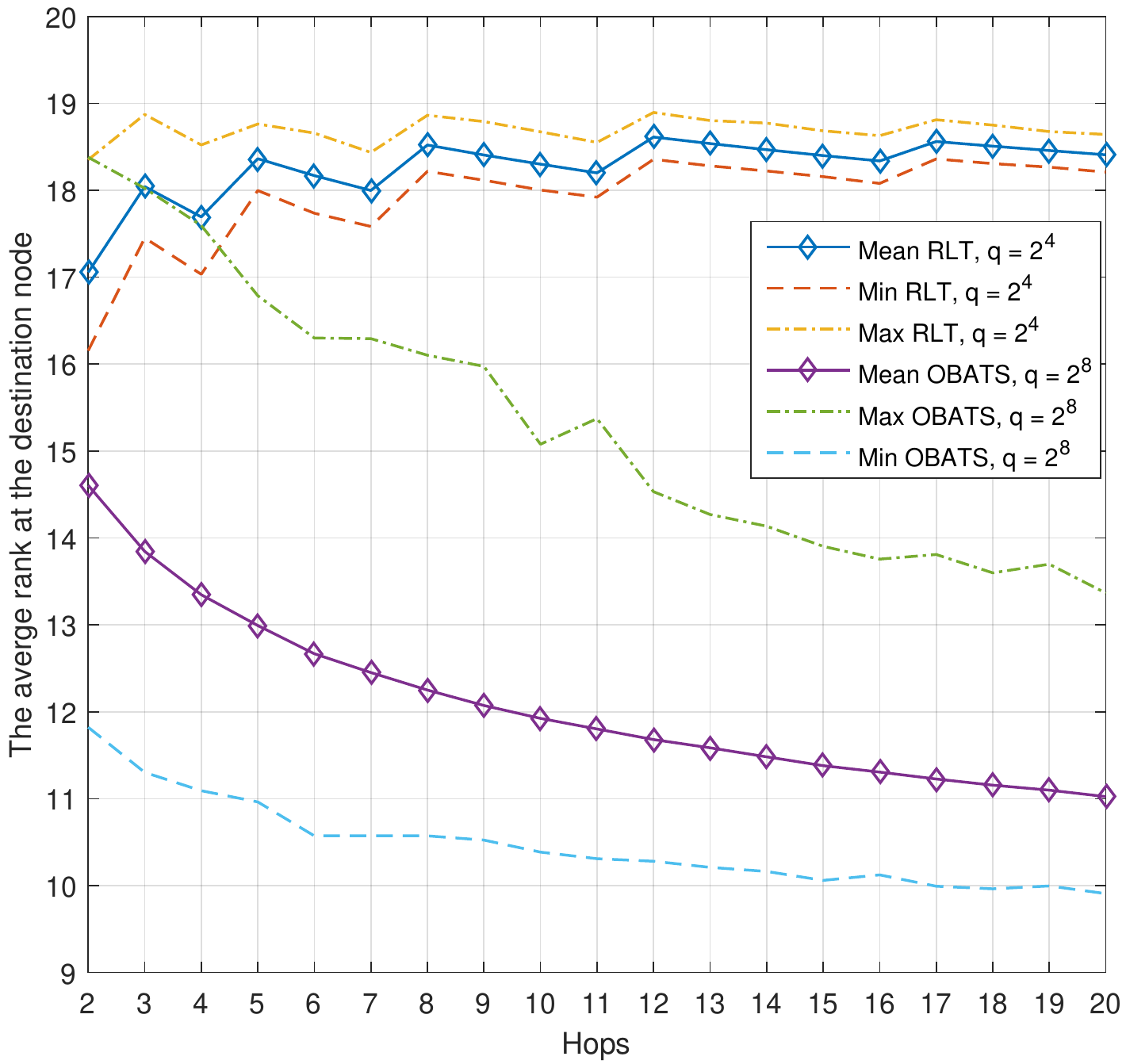}
\label{avg_rank_M20}}
\hfil
\subfloat[$M = 24$]{\includegraphics[width=3in]{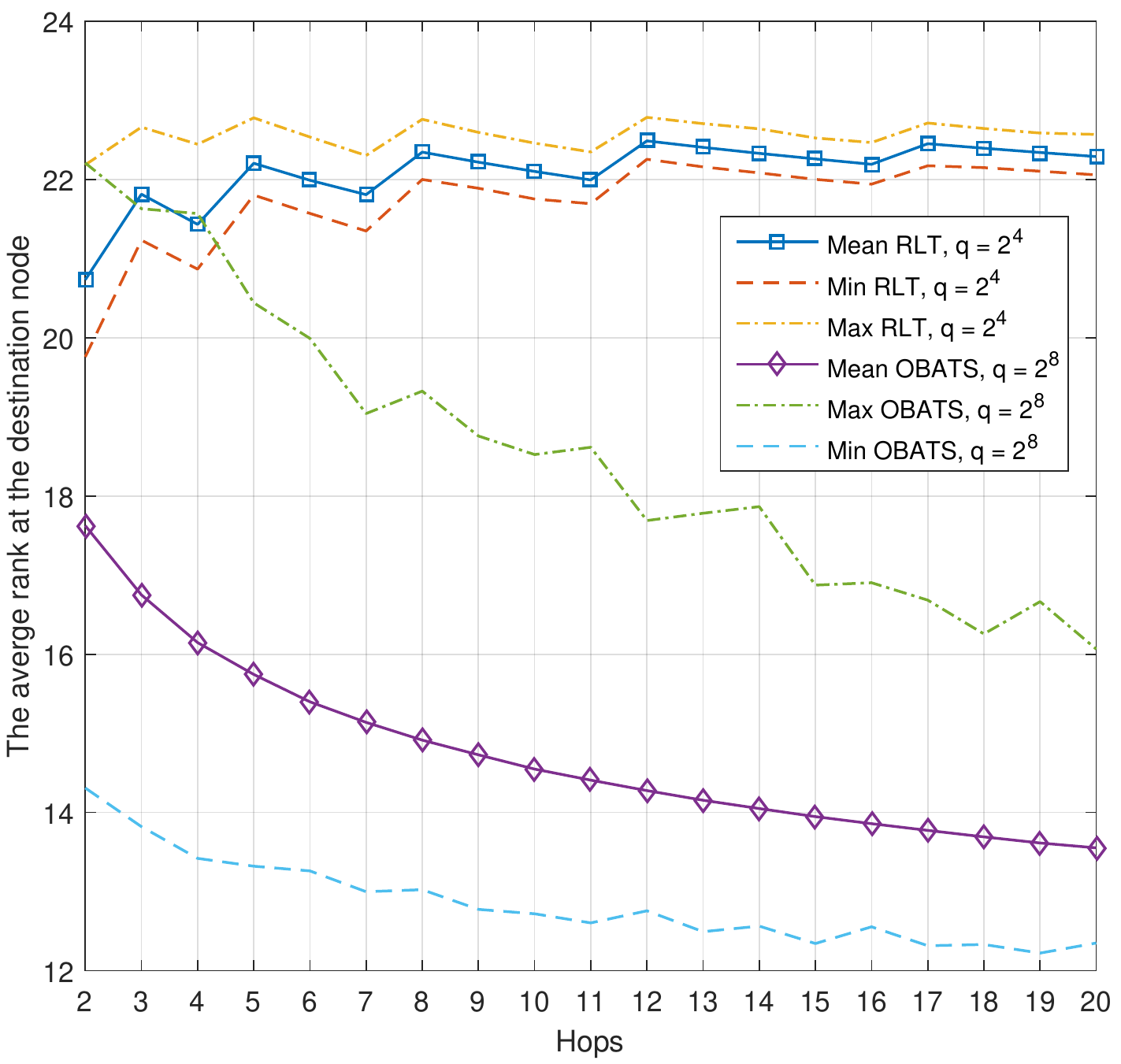}
\label{avg_rank_M24}}
\hfil
\caption{The average rank vs the number of hops under different batch size.}
\label{avg_rank}
\end{figure*}

In this subsection, we investigate another important characteristic of BATS codes, i.e., the average rank of transfer matrix at the destination node, which determines that the least number of batches need be generated at the source. For the purpose of brevity, we only compare the RLT-based algorithm\footnote{The average ranks between the RLT-based and CLT-based algorithm are very similar.} with the original BATS, and set the field size $q = 2^4$ for the former while $q = 2^8$ for the latter.\footnote{We also observe that the average ranks of the RLT-based algorithm are very close when $q = 2^4$ and $q = 2^8$. Therefore, only the values corresponding to $q = 2^4$ are plotted due to the lower complexity.}

The first thing to note is that the average rank of our proposed approach is close to the batch size, and the change trend goes up with the length of path. In the meanwhile, OBATS leads to smaller average ranks that decrease with the number of hops. For the purpose of illustration, the dash and dash-dot lines that represent the smallest and the largest values in the sample, respectively, are also plotted in Fig. \ref{avg_rank}. Obviously, the fluctuation of the average rank generated by our proposed approach are much smoother than that by OBATS.

In practical systems, the question of ``When to stop generating batches at sources'', which critically depends on the rank distribution, is crucial to the implementation of BATS codes. It is very difficult to solve this problem due to the complex environment. For example, since there exist multiple paths in multicast networks, the number of batches generated by the source is determined by the path with the smallest average rank at destination node. Hence, if the average ranks at sink nodes are far apart from each other, the performance for the pathes with large average rank may suffer significant loss. The reason is that intermediate nodes belonging to such pathes will process much more batches than expected.

The results in Fig. \ref{avg_rank} suggest us a potential way of answering the question mentioned above. Since the average ranks by our approximating methods are very close, the numbers of batches related to different network conditions are likely to be close. However, we should note that, there is still, unfortunately, a considerable gap between the minimum and maximum ranks. Therefore, how to bridge the gap needs to be investigated further with other complementary methods, such as to take into account a tradeoff between the number of batches and the number of coded packets of a batch at intermediate nodes.  More detailed descriptions of this issue will be addressed in our future work.

\begin{figure}[!t]
\centering
\includegraphics[width=3.2in]{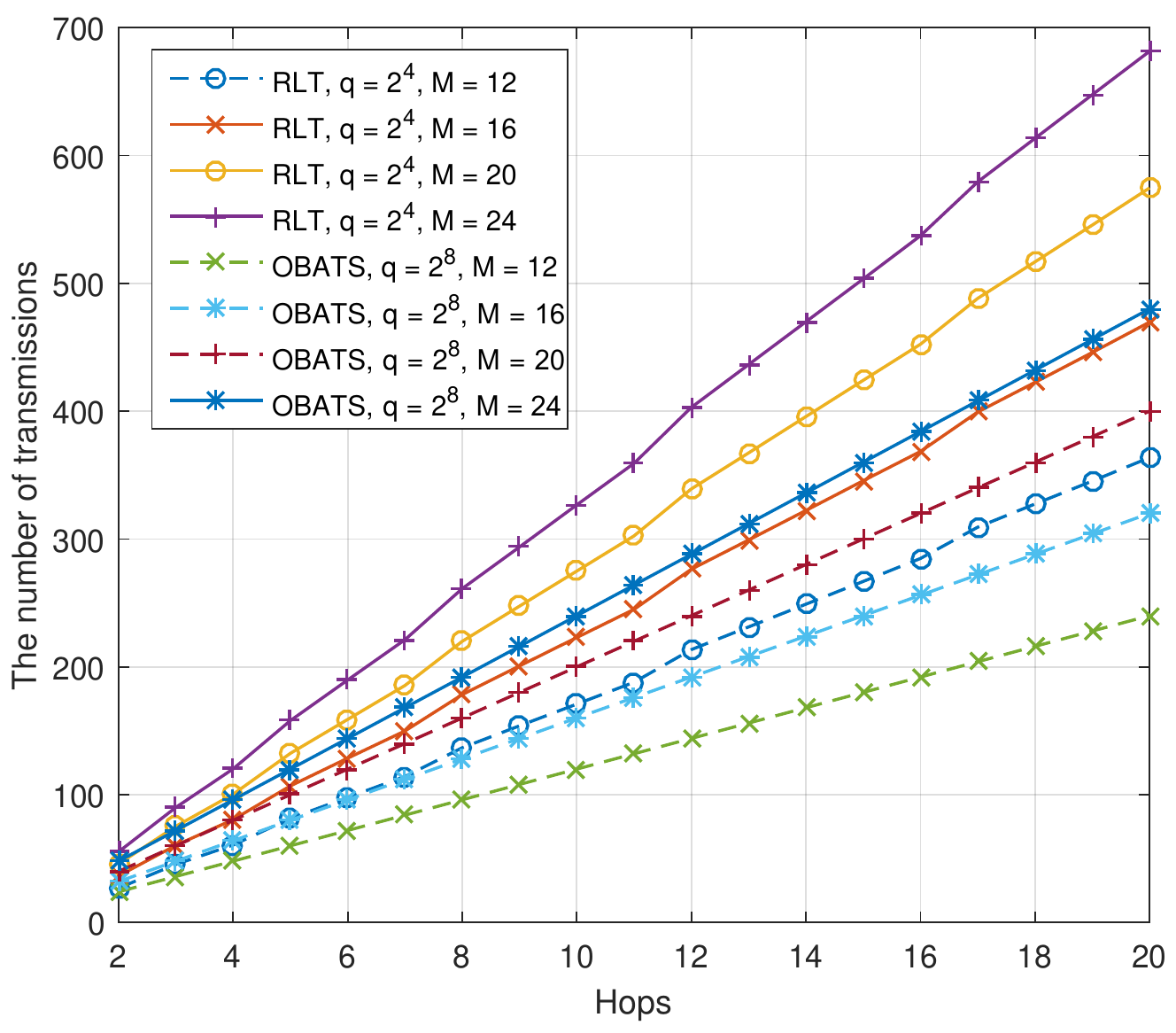}
\caption{The total number of transmissions per batch vs the number of hops under different batch size.}
\label{total_number_trans}
\end{figure}

\subsection{The number of transmission}

Finally, Fig. \ref{total_number_trans} provides the overall total number of transmission required to deliver a batch from the source node to the destination node, i.e., $t_{total} = \sum_k^l t_k$,. It can be observed that, though the RLT-based algorithm requires more transmissions than the original method with the same batch size $M$, our approach provide higher transmission efficiency. It is worth to point out that even if the original BATS code uses a greater $M$, our approach still has a good performance in terms of both the transmission efficiency and average rank. This fact indicates that making coding decision with current condition can lead to better outcomes.

\section{Conclusion}
In this work, we conducted an optimization framework for evaluating the lower bound on the number of packets from the source node to the sink node over multi-hop wireless networks. The framework relied on the relation between the number of transmissions and the rank distribution of the received batches, and was represented as an MINLP problem. By exploiting the properties of the MINLP formulation, we developed the explicit expression for computing the rank distribution. In addition, an NLP formulation was proposed as the upper bound on our problem, which was also used as the performance measurement. Using these properties again, the global and local real-time approaches are designed. Finally, we presented the numerical experiments to illustrate the performance of the proposed approaches in terms of transmission efficiency and average rank. The interesting question that remains open is how the algorithm described in this paper can be extended to multicast networks.



\ifCLASSOPTIONcaptionsoff
  \newpage
\fi

\appendices
\counterwithin{lemma}{section}\section{Proof of Thoerem \ref{th:rank_dist}}\label{app-A}
For the rank of a totally random matrix, we have the following lemma.
\begin{lemma}
\label{le:A.1}
\textit{
The probability that the rank of a totally random matrix $\mathbf{M} \in \mathbb{F}_q^{n \times m}$, denoted by rk($\mathbf{M}$), is $r \le {\rm{min}}(s,t) $ is given by~\cite{Fulman15}
}
\end{lemma}
\begin{align}
\label{eq:rank.of.random.mtrx}
\Pr \{ \text{rk} (\mathbf{M}) = r \} & = \frac{1}{q^{(n - r)(m - r)}} \prod\limits_{k=0}^{r-1} {\frac{(1 - q^{-n+k}) (1 - q^{-m+k})}{(1 - q^{-r+k})}} \noindent \\
& = \zeta_{r}^{n,m}.
\end{align}

Another lemma that will be useful is as follows.
\begin{lemma}
\label{le:A.2}
\textit{
Consider an invertible matrix $\mathbf{M} \in \mathbb{F}_q^{n \times n}$ and a random vector $\mathbf{v} \in \mathbb{F}_q^n$ with uniform, independent entries. Let $\mathbf{w} = \mathbf{M} \mathbf{v}$. Then, vector $\mathbf{w}$ is uniformly distributed over $\mathbb{F}_q^n$.
}
\end{lemma}

\begin{IEEEproof}
As $\mathbf{M}$ is invertible, we can write $\mathbf{M} = \mathbf{I} \cdot
\prod_{i=1}^{k} \mathbf{E}_i$, $k \ge 1$, where $\mathbf{I}$ is an identity matrix and $\mathbf{E}_i$ are elementary matrices. First, we claim that $\mathbf{\hat{w}} = \mathbf{E}_i \mathbf{v}$ is uniformly distributed over $\mathbb{F}_q^n$. When matrix $\mathbf{E}_i$ corresponds to switching or multiplication operation, the claim is clearly true. Next, let us consider $\mathbf{E}_i$ with an element, called $m \in \mathbb{F}_q$, in the ($i,j$) position. Without loss of generality, let $i = 1$, $j = 2$, and $\mathbf{v} = [v_1,\, v_2,\,  \cdots,\, v_n]$. Thus, $\mathbf{\hat{w}} = \mathbf{E}_i \mathbf{v} = [v_1 + m v_2,\, v_2,\,  \cdots,\, v_n]$. According to the assumption, $v_2, \ldots, v_n$ are i.i.d. Then, let random variable $\hat{v} = v_1 + m v_2$. We have
\begin{equation}
\Pr (\hat{v} = x) = \sum\limits_{i \in \mathbb{F}_q} \Pr (v_2 = i) \Pr (v_1 = x - m i) = \frac{1}{q}, \quad x \in \mathbb{F}_q. \nonumber
\end{equation}

That is, $\hat{v}$ follows uniform distribution. Also, we write
\begin{align}
\Pr (\hat{v} = x, v_2 = y)
&= \Pr (\hat{v} = x | v_2 = y) \Pr (v_2 = y) \nonumber \\
&= \Pr (v_1 = x - m y) \Pr (v_2 = y) \nonumber \\
&= \frac{1}{q^2} \nonumber \\
& = \Pr (\hat{v} = x) \Pr (v_2 = y), \quad x,y \in \mathbb{F}_q. \nonumber
\end{align}

That is, $\hat{v}$ is independent of $v_2$. Therefore, it implies that $\mathbf{\hat{w}}$ is uniformly distributed over $\mathbb{F}_q^n$. Finally, since $\mathbf{\hat{w}} = \mathbf{I} \mathbf{\hat{w}}$, the proof is completed.
\end{IEEEproof}

With the above lemmas, we then get

\begin{lemma}
\label{le:A:3}
\textit{
Let $\mathbf{A} \in \mathbb{F}_q^{s \times t}$ be a random matrix with arbitrary probability distribution, and let $\mathbf{B} \in \mathbb{F}_q^{t \times m}$ be a totally random matrix. The probability that the rank of matrix $\mathbf{A} \mathbf{B}$, conditional on $ {\rm{rk}} (\mathbf{A}) = n $, is $r \le {\rm{min}}(n,m,t) $ is given by
}
\end{lemma}

\begin{equation}
\label{eq:rank.of.prod.of.mtrx}
\Pr \{ \text{rk} (\mathbf{AB}) = r | \text{rk} (\mathbf{A}) = n \} = \zeta_{r}^{n,m}.
\end{equation}

\begin{IEEEproof}
Applying the elementary row and column operation to $\mathbf{A}$, we get
\begin{equation}
\mathbf{A} = \mathbf{L}
\begin{bmatrix} \mathbf{I}_A & 0 \\ 0 & 0 \end{bmatrix}
\mathbf{U}, \nonumber
\end{equation}

\noindent where $\mathbf{L}$ and $\mathbf{U}$ are invertible matrices, and $\mathbf{I}_A$ is a $\text{rk} (\mathbf{A}) \times \text{rk} (\mathbf{A})$ identity matrix. We write
\begin{align*}
&\Pr \left\{ \text{rk} (\mathbf{AB}) \, | \, \text{rk} (\mathbf{A}) \right\}\\
& = \Pr \left\{ \text{rk} \left(\mathbf{L} \begin{bmatrix} \mathbf{I}_A & 0 \\ 0 & 0 \end{bmatrix} \mathbf{U B} \right) \, \middle| \, \text{rk} (\mathbf{I}_A) \right\} \\
& = \Pr \left\{ \text{rk} \left(\begin{bmatrix} \mathbf{I}_A & 0 \\ 0 & 0 \end{bmatrix} \mathbf{U B}\right) \, \middle| \, \text{rk} (\mathbf{I}_A) \right\} \\
& = \Pr \left\{ \text{rk} (\mathbf{C}) \right\},
\end{align*}

\noindent where $\mathbf{C}$ is an $n \times m$ matrix obtained by keeping the first $i$ rows of $\mathbf{U B}$. According to Lemma \ref{le:A.2}, $\mathbf{U B}$ is a totally random matrix, so is $\mathbf{C}$. Then, the proof is directly completed by Lemma \ref{le:A.1}.
\end{IEEEproof}

In this paper, the transfer matrix can be fully described as follows. With slight abuse of notation, we use $\mathbf{H}_{k}$ to represent the transfer matrix obtained by node $v_{k}, k \ge 1$\footnote{Let $\mathbf{H}_{k,i}$ be the transfer matrix corresponding to the $i$-th batch received by node $v_{k}, k \ge 1$. Here, we ignore the subscript $i$, because each batch is independently encoded and retransmitted such that the matrices $\mathbf{H}_{k,i}$ obtained by node $v_k$ are i.i.d.}. Define $\mathbf{\Phi}_{k}$ as the $t_{k-1} \times t_{k}$ totally random matrix generated by node $v_k$. In particular, $t_0 = M$. Define $\mathbf{D}_{k} $ as a $t_{k} \times t_{k}$ random diagonal matrix consisting of independent diagonal entries $d_{jj} = 1$ with probability $1 - \epsilon_k$ and $d_{jj} = 0$ with probability $\epsilon_k, j = 1, 2,\ldots, t_k$. The transfer matrix $\mathbf{H}_{k+1}$ can, then, be expressed as,
\begin{equation}
\label{eq:trans.matrix}
\mathbf{H}_{k+1} = \mathbf{H}_{k}\mathbf{\Phi}_{k}\mathbf{D}_{k}, \quad k = 1, \ldots, l,
\end{equation}

\noindent where $\mathbf{H}_1 = \text{diag}(1, 1,\ldots, 1)$ is an $M \times M$ identity matrix. That is, the rank of $\mathbf{H}_{k} \, (1 \le k \le l)$ is no more than $M$. In addition, we postulate that $\mathbf{\Phi}_{1}, \ldots, \mathbf{\Phi}_{l}, \mathbf{D}_{1}, \ldots, \mathbf{D}_{l}$ are mutually independent.

\begin{IEEEproof}[Proof of Theorem \ref{th:rank_dist}]
Let index set $\Omega_k = \{i: \mathbf{d}_{k,i} \neq \mathbf{0}, i = 1,2, \ldots, t_k \}$, where $\mathbf{d}_{k,i}$ denotes the $i$-th column of $\mathbf{D}_{k}$. Using Eq. (\ref{eq:trans.matrix}), we have
\begin{align}
\label{eq:th2.1}
h_{k+1,r}
=& \Pr \{\text {rk}(\mathbf{H}_{k+1}) = r \} = \Pr \{\text{rk}(\mathbf{H}_{k}\mathbf{\Phi}_{k}\mathbf{D}_{k}) = r\} \nonumber \\
=& \sum\limits_{m=r}^{M} {\sum\limits_{n=r}^{t_k} {\Pr \{ \text{rk}(\mathbf{H}_{k}\mathbf{\Phi}_{k}\mathbf{D}_{k}) = r, \, \text{rk}(\mathbf{H}_{k}) = m, \, |\Omega_k| = n\} }} \nonumber \\
=& \sum\limits_{m=r}^{M} {\sum\limits_{n=r}^{t_k} {\Pr \{\text{rk}(\mathbf{H}_{k}) = m\} \Pr \{|\Omega_k| = n\}}} \Pr \{\text{rk}(\mathbf{H}_{k}\mathbf{\Phi}_{k}\mathbf{D}_{k}) = r \, | \, \text{rk}(\mathbf{H}_{k}) = m, \, |\Omega_k| = n\},
\end{align}

\noindent where $\Pr \{\text{rk}(\mathbf{H}_{k}) = m\} = h_{k,m}$, and  the probability of the cardinality of set $\Omega_k$ follows a binomial distribution, i.e., $\Pr \{ |\Omega_k| = n \} = \binom{t_{k}}{n} (1 - \epsilon_k)^n \, \epsilon_k^{t_{k} - n}$. Moreover, the cardinality of set $\Omega_k$ implies that $\mathbf{\Phi}_{k}\mathbf{D}_{k} \in \mathbb{F}_q^{t_{k-1} \times |\Omega_k|}$. Consequently, the proof is completed by applying Lemma \ref{le:A:3}.
\end{IEEEproof}
\section{Proof of Lemma \ref{le:1}}\label{app-B}
\begin{IEEEproof}[Proof of Lemma \ref{le:1}]
Since matrix $\mathbf{P}_k$ is a lower-triangular matrix, its diagonal components are the eigenvalues. Let $\mathbf{Q} = [\mathbf{q}_1, \mathbf{q}_2, \ldots, \mathbf{q}_{M+1}]$. To complete the proof, we need to exam the following equality
\begin{equation}
\label{eq:le3.1}
(\mathbf{P}_k - \lambda_{k,j} \mathbf{I}) \mathbf{q}_j = \mathbf{0}.
\end{equation}

\noindent Case 1: $j = 1$.

As $\mathbf{P}_k$ is a translation matrix, we have
\begin{align}
\label{eq:le3.2}
\lambda_{k,1} & = 1 = \sum_{i = 1}^{M + 1} \sum_{n=i - 1}^{t_k} f(k,n) \zeta_{i-1}^{m-1,n}.
\end{align}

Substituting (\ref{eq:le3.2}) into (\ref{eq:le3.1}), it is clear that the claim holds for $j = 1$.

\noindent Case 2: $j= 2, 3, \ldots, M + 1$.

Let us consider $\lambda_{k,j}$ and the $(m+1)$-th component $\mathbf{\alpha}_{m+1}$ of vecter $(\mathbf{P}_k - \lambda_{k,j} \mathbf{I}) \mathbf{q}_{j}$, where
\begin{align}
\label{eq:le3.alpha}
\mathbf{\alpha}_{m+1}
=& \underbrace{\sum_{n=j-1}^{t_k} f(k,n) \zeta_{j-1}^{m,n} \zeta_{j-1}^{j-1}}_{(1)} + \underbrace{\sum_{n=j}^{t_k} f(k,n) \zeta_{j}^{m,n} \zeta_{j-1}^{j}}_{(2)} + \cdots \nonumber \\
& + \underbrace{\sum_{n=m-1}^{t_k} f(k,n) \zeta_{m-1}^{m,n} \zeta_{j-1}^{m-1}}_{(m-j+1)} + \underbrace{\sum_{n=m}^{t_k} f(k,n) \zeta_{m}^{m,n} \zeta_{j-1}^{m}}_{(m-j+2)} \nonumber \\
& - \underbrace{\sum_{n=j-1}^{t_k} f(k,n) \zeta_{j-1}^{j-1,n} \zeta_{j-1}^{m}}_{(0)}, \quad 3 \le j \le M + 1, \; j \le m \le M + 1.
\end{align}

\noindent In particular, $\mathbf{\alpha}_{m+1} = 0, \, m < j$. We expand each component of Eq. (\ref{eq:le3.alpha}) into Eq. (\ref{eq:le3.expansion}).
\begin{table*}
\begin{equation}
\label{eq:le3.expansion}
\setlength{\arraycolsep}{0pt}
\begin{array}{cccccccccc}
(0) & &&&&&&&&\\
  =&  f(k,j-1) \zeta_{j-1}^{j-1,j-1} \zeta_{j-1}^{m} &+& f(k,j) \zeta_{j-1}^{j-1,j} \zeta_{j-1}^{m} &+& \cdots &+& f(k,m-1) \zeta_{j-1}^{j-1,m-1} \zeta_{j-1}^{m} &+& \sum\limits_{n=m}^{t_k} f(k,n) \zeta_{j-1}^{j-1,n} \zeta_{j-1}^{m}, \\
(1) & &&&&&&&&\\=&  f(k,j-1) \zeta_{j-1}^{m,j-1} \zeta_{j-1}^{j-1} &+& f(k,j) \zeta_{j-1}^{m,j} \zeta_{j-1}^{j-1} &+& \cdots &+& f(k,m-1) \zeta_{j-1}^{m,m-1} \zeta_{j-1}^{j-1} &+& \sum\limits_{n=m}^{t_k} f(k,n) \zeta_{j-1}^{m,n} \zeta_{j-1}^{j-1}, \\
(2) & &&&&&&&&\\ =&  0 &+& f(k,j) \zeta_{j}^{m,j} \zeta_{j-1}^{j} &+& \cdots &+& f(k,m-1) \zeta_{j}^{m,m-1} \zeta_{j-1}^{j} &+& \sum\limits_{n=m}^{t_k} f(k,n) \zeta_{j}^{m,n} \zeta_{j-1}^{j}, \\
 & \vdots & \vdots & \vdots &\\
(m-j+1) & &&&&&&&&\\ =&  0 &+& 0 &+& \cdots &+& f(k,m-1) \zeta_{m-1}^{m,m-1} \zeta_{j-1}^{m-1} &+& \sum\limits_{n=m}^{t_k} f(k,n) \zeta_{m-1}^{m,n} \zeta_{j-1}^{m-1}, \\
(m-j+2) & &&&&&&&&\\ =&  0 &+& 0 &+& \cdots &+& 0 &+& \sum\limits_{n=m}^{t_k} f(k,n) \zeta_{m}^{m,n} \zeta_{j-1}^{m}.
\end{array}
\end{equation}
\hrulefill
\end{table*}

Replacing (\ref{eq:le3.expansion}) back into (\ref{eq:le3.alpha}), we get
\begin{align}
\label{eq:le3.alpha.expansion}
\mathbf{\alpha}_{m+1}
=& f(k,j-1) \left( \zeta_{j-1}^{m,j-1} \zeta_{j-1}^{j-1} - \zeta_{j-1}^{j-1,j-1} \zeta_{j-1}^{m} \right) + f(k,j) \left( \sum_{k=0}^{1} \zeta_{j-1+k}^{m,j} \zeta_{j-1}^{j-1+k} - \zeta_{j-1}^{j-1,j} \zeta_{j-1}^{m} \right) \nonumber \\
& + \cdots + f(k,m-1) \left( \sum_{k=0}^{m-j} \zeta_{j-1+k}^{m,m-1} \zeta_{j-1}^{j-1+k} - \zeta_{j-1}^{j-1,m-1} \zeta_{j-1}^{m} \right) \nonumber \\
& + \sum_{n=m}^{t_k} f(k,n) \underbrace{\left(\sum_{k=0}^{m-j+1} \zeta_{j-1+k}^{m,n} \zeta_{j-1}^{j-1+k} - \zeta_{j-1}^{j-1,n} \zeta_{j-1}^{m} \right)}_{(a)}.
\end{align}

Next, we rewrite equation (a) to
\begin{align}
\label{eq:le3.a}
&\sum_{k=0}^{m-j+1} \zeta_{j-1+k}^{m,n} \zeta_{j-1}^{j-1+k} - \zeta_{j-1}^{j-1,n} \zeta_{j-1}^{m} \nonumber \\
& = \sum_{k=0}^{m-j+1} \frac{\zeta_{j-1+k}^{m} \zeta_{j-1+k}^{n} \zeta_{j-1}^{j-1+k}}{\zeta_{j-1+k}^{j-1+k} q^{(m-j+1-k)(n-j+1-k)}} - \frac{\zeta_{j-1}^{j-1} \zeta_{j-1}^{n} \zeta_{j-1}^{m}}{\zeta_{j-1}^{j-1}} \nonumber \\
& \overset{(b)}{=} \zeta_{j-1}^{m} \zeta_{j-1}^{n} \left( \sum_{k=0}^{m-j+1} \frac{\zeta_{k}^{m-j+1} \zeta_{k}^{n-j+1}}{\zeta_{k}^{k} q^{[(m-j+1)-k][(n-j+1)-k]}} - 1 \right) \nonumber \\
& = \zeta_{j-1}^{m} \zeta_{j-1}^{n} \left( \sum_{k=0}^{m-j+1} \zeta_{k}^{m-j+1,n-j+1} - 1 \right),
\end{align}

\noindent where (b) follows the fact that
\begin{align*}
\zeta_{j-1+k}^{m}
& = \prod_{r = 0}^{j + k - 2} (1 - q^{-m+ r}) \\
& = \prod_{x=0}^{j-2} (1 - q^{-m+x}) \prod_{y=0}^{k-1} (1 - q^{-(m-j+1) + y}) \nonumber \\
& = \zeta_{j-1}^{m} \zeta_{k}^{m-j+1},
\end{align*}

Applying (\ref{eq:le3.a}) to (\ref{eq:le3.alpha.expansion}), we get
\begin{align*}
\mathbf{\alpha}_{m+1}
=& f(k,j-1) \zeta_{j-1}^{m} \zeta_{j-1}^{j-1} \left( 1 - 1 \right)
+ f(k,j) \zeta_{j-1}^{m} \zeta_{j-1}^{j} \left( \sum_{k=0}^{1} \zeta_{k}^{m-j+1,1} - 1 \right) \\
& + \cdots + f(k,m-1) \zeta_{j-1}^{m} \zeta_{j-1}^{m-1} \left( \sum_{k=0}^{m-j} \zeta_{k}^{m-j+1,m-j} - 1 \right) \\
& + \sum_{n=m}^{t_k} f(k,n) \zeta_{j-1}^{m} \zeta_{j-1}^{n} \left(\sum_{k=0}^{m-j+1} \zeta_{k}^{m-j+1,n-j+1} - 1 \right)  \\
\overset{(c)}{=}& f(k,j-1) \zeta_{j-1}^{m} \zeta_{j-1}^{j-1} \times 0 + f(k,j) \zeta_{j-1}^{m} \zeta_{j-1}^{j} \times 0 + \cdots \\
& + f(k,m-1) \zeta_{j-1}^{m} \zeta_{j-1}^{m-1} \times 0 + \sum_{n=m}^{t_k} f(k,n) \zeta_{j-1}^{m} \zeta_{j-1}^{n} \times 0 \\
=& 0,
\end{align*}

\noindent where (c) is derived from Lemma \ref{le:A.1},
\begin{align*}
\sum_{k=0}^{m-j+1} \zeta_{k}^{m-j+1,n-j+1} = \sum_{k=0}^{m-j+1} \Pr \left\{ \text{rk}(\mathbf{M}) = k \, | \, \mathbf{M} \in \mathbb{F}_q^{(m-j+1) \times (n-j+1)} \right\} = 1, \quad j \le m \le n.
\end{align*}

The proof is completed.
\end{IEEEproof}

\section{Proof of Theorem \ref{th:5}}\label{app-C}

In order to prove Theorem \ref{th:5}, we need the following lemmas.

\begin{lemma}
\label{le:C1}
\textit{
Let $\mathbf{\Phi} \in \mathbb{F}_q^{m \times n}$ and $ \mathbf{\hat{\Phi}} \in \mathbb{F}_{\hat{q}}^{m \times n}$ are totally random matrices. If $q < \hat{q}$, then $\Pr\{\rm{{rk}}(\mathbf{\Phi}) \ge r\} \le \Pr\{\rm{{rk}}(\mathbf{\hat{\Phi}}) \ge r\}, \; 0 \le r \le \min(m,n)$. The equality holds if and only if $r = 0$.
}
\end{lemma}

\begin{IEEEproof}
Without loss of generality, we assume $n \le m$. Then, the theorem is proved by induction on $n$. For $n = 1, 2$, it can be verified by the following facts,
\begin{align*}
\Pr\{\text{rk}(\mathbf{\Phi}) = 0\} & = q^{-mn} > \hat{q}^{-mn} = \Pr\{\text{rk}(\mathbf{\hat{\Phi}}) = 0\}, \\
\Pr\{\text{rk}(\mathbf{\Phi}) = n\} & = \prod_{i=0}^{n-1} (1 - q^{-m+i}) < \prod_{i=0}^{n-1} (1 - \hat{q}^{-m+i}) \\
& = \Pr\{\text{rk}(\mathbf{\hat{\Phi}}) = n\}.
\end{align*}

Suppose the claim holds for $n = l - 1$. Then, let us consider $n = l$. In particular, we define the $m \times l$ matrices as $\mathbf{\Phi}^l = [\mathbf{a}_1,\ldots,\mathbf{a}_l]$ and $\mathbf{\hat{\Phi}}^{l} = [\mathbf{\hat{a}}_1,\ldots,\mathbf{\hat{a}}_l]$, and let $\langle \mathbf{\Phi}^l \rangle$ be the column space of $\mathbf{\Phi}^l$. We have
\begin{align}
\label{eq:leC1.1}
\Pr \{ \text{rk} (\mathbf{\Phi}^{l}) \ge r \}
= & \Pr \{ \text{rk} (\mathbf{\Phi}^{l-1}) \ge r \} + \Pr \{ \text{rk} (\mathbf{\Phi}^{l-1}) = r - 1, \mathbf{a}_{l} \notin \langle \mathbf{\Phi}^{l-1} \rangle \} \nonumber \\
= & \Pr \{ \text{rk} (\mathbf{\Phi}^{l-1}) \ge r \} + (1 - q^{-m+r-1}) \Pr \{ \text{rk} (\mathbf{\Phi}^{l-1}) = r - 1 \}, \\
\label{eq:leC1.2}
\Pr \{ \text{rk} (\mathbf{\hat{\Phi}}^{l}) \ge r \}
= & \Pr \{ \text{rk} (\mathbf{\hat{\Phi}}^{l-1}) \ge r \} + (1 - \hat{q}^{-m+r-1}) \Pr \{ \text{rk} (\mathbf{\hat{\Phi}}^{l-1}) = r - 1 \}, r > 0
\end{align}

\noindent where $\mathbf{\Phi}^{l-1} = [\mathbf{a}_1,\ldots,\mathbf{a}_{l-1}]$ and $\mathbf{\hat{\Phi}}^{l-1} = [\mathbf{\hat{a}}_1,\ldots,\mathbf{\hat{a}}_{l-1}]$ are submatrices of $\mathbf{\Phi}^{l}$ and $\mathbf{\hat{\Phi}}^{l}$, respectively. To compare Eq. (\ref{eq:leC1.1}) and (\ref{eq:leC1.2}), there are two cases:

Case 1: $\Pr \{ \text{rk} (\mathbf{\Phi}^{l-1}) = r - 1 \} < \Pr \{ \text{rk} (\mathbf{\hat{\Phi}}^{l-1}) = r - 1 \}$. With the assumption and $(1 - q^{-m+r-1}) < (1 - \hat{q}^{-m+r-1})$, we get $\Pr\{\text{rk}(\mathbf{\Phi}^l) \ge r\} < \Pr\{\text{rk}(\mathbf{\hat{\Phi}}^l) \ge r \}$.

Case 2: $\Pr \{ \text{rk} (\mathbf{\Phi}^{l-1}) = r - 1 \} \ge \Pr \{ \text{rk} (\mathbf{\hat{\Phi}}^{l-1}) = r - 1 \}$. We rewrite Eq. (\ref{eq:leC1.1}) and (\ref{eq:leC1.2}) into
\begin{align*}
\Pr \{ \text{rk} (\mathbf{\Phi}^{l}) \ge r \}
= & \Pr \{ \text{rk} (\mathbf{\Phi}^{l-1}) \ge r - 1 \} - q^{-m+r-1} \Pr \{ \text{rk} (\mathbf{\Phi}^{l-1}) = r - 1 \}, \\
\Pr \{ \text{rk} (\mathbf{\hat{\Phi}}^{l}) \ge r \}
= & \Pr \{ \text{rk} (\mathbf{\hat{\Phi}}^{l-1}) \ge r - 1 \} - \hat{q}^{-m+r-1} \Pr \{ \text{rk} (\mathbf{\hat{\Phi}}^{l-1}) = r - 1 \}.
\end{align*}

\noindent With the assumption and $q^{-m+r-1} > \hat{q}^{-m+r-1}$, we obtain $\Pr\{\text{rk}(\mathbf{\Phi}^l) \ge r\} < \Pr\{\text{rk}(\mathbf{\hat{\Phi}}^l) \ge r \}$.

In particular, $\Pr\{\text{rk}(\mathbf{\Phi}^l) \ge 0\} = \Pr\{\text{rk}(\mathbf{\hat{\Phi}}^l) \ge 0 \} = 1$. Consequently, the proof is completed by induction.
\end{IEEEproof}

\begin{lemma}
\label{le:C2}
\textit{
If $\mathbf{\Phi} \in \mathbb{F}_q^{m \times n}$ and $ \mathbf{\hat{\Phi}} \in \mathbb{F}_{q}^{m \times \hat{n}}, n < \hat{n}$, are totally random matrices, then $\Pr\{\rm{{rk}}(\mathbf{\Phi}) \ge r\} \le \Pr\{\rm{{rk}}(\mathbf{\hat{\Phi}}) \ge r\}, \; 0 \le r \le \min(m,\hat{n})$. The equality holds if and only if $r = 0$.
}
\end{lemma}

The proof is similar to that of Lemma \ref{le:C1} and thus is omitted.

\begin{lemma}
\label{le:C3}
\textit{
The average rank increases with finite field size $q$ with given $t_k$, k = 1, \ldots, l.
}
\end{lemma}

\begin{IEEEproof}
Suppose $q < \hat{q}$, and the elements of totally random matrices $\mathbf{\Phi}_k$ and $\mathbf{\hat{\Phi}}_k$, $1 \le k \le l$, are chosen from $\mathbb{F}_q$ and $\mathbb{F}_{\hat{q}}$, respectively. We claim that $\Pr \{ \text{rk} (\mathbf{H}_{k+1} = \mathbf{H}_k \mathbf{\Phi}_k \mathbf{D}_k) \ge r\} < \Pr \{ \text{rk} (\mathbf{\hat{H}}_{k+1} = \mathbf{\hat{H}}_k \mathbf{\hat{\Phi}}_k \mathbf{D}_k) \ge r \}, r > 0$. This can be proved by induction on $k$, the number of hops. For $k = 1$, we have
\begin{align}
&\Pr \{ \text{rk} (\mathbf{H}_2 = \mathbf{H}_1 \mathbf{\Phi}_1 \mathbf{D}_1) \ge r \} = \Pr \{ \text{rk} (\mathbf{\Phi}_1 \mathbf{D}_1) \ge r \} \nonumber \\
\overset{(a)}{<} &\Pr \{ \text{rk} (\mathbf{\hat{H}}_2 = \mathbf{H}_1 \mathbf{\hat{\Phi}}_1 \mathbf{D}_1) \ge r \} = \Pr \{ \text{rk} (\mathbf{\hat{\Phi}}_1 \mathbf{D}_1) \ge r \} , \quad r > 0. \nonumber
\end{align}

\noindent where (a) follows by Lemma \ref{le:C1}, as $\mathbf{\Phi}_1 \mathbf{D}_1$ and $\mathbf{\hat{\Phi}}_1 \mathbf{D}_1$ belong to the subspaces with the same dimension.

Next, suppose that the claim holds for $k = l - 1$. Then, let us consider $k = l$, and let random matrix $\mathbf{\bar{H}}_l \in \mathbb{F}_{\hat{q}}^{M \times t_k}$ have the same rank distribution as $\mathbf{H}_l$, i.e. $\Pr \{ \text{rk} (\mathbf{\bar{H}}_{l}) \ge r\} = \Pr \{ \text{rk} (\mathbf{H}_{l}) \ge r \}, r > 0$. We write
\begin{align*}
\Pr \{ \text{rk} (\mathbf{\hat{H}}_{l+1} = \mathbf{\hat{H}}_l \mathbf{\hat{\Phi}}_l \mathbf{D}_l) \ge r \}
& = \sum_{x=r}^{M} \Pr\{ \text{rk} (\mathbf{\hat{H}}_l \mathbf{\hat{\Phi}}_l \mathbf{D}_l) = x \} \\
& = \sum_{x=r}^{M} \sum_{m = x}^{M} \Pr\{ \text{rk} (\mathbf{\hat{H}}_l) = m \} g(m,x,l,\hat{q}) \\
& = \sum_{m=r}^{M} \hat{h}_{l,m} \left( \sum_{x=r}^{m} g(m,x,l,\hat{q}) \right) \\
& \overset{(b)}{>} \sum_{m=r}^{M} \bar{h}_{l,m} \left( \sum_{x=r}^{m} g(m,x,l,\hat{q}) \right) \\
& \overset{(c)}{>} \sum_{m=r}^{M} h_{l,m} \left( \sum_{x=r}^{m} g(m,x,l,q) \right) \\
& = \Pr \{ \text{rk} (\mathbf{H}_{l+1} = \mathbf{H}_l \mathbf{\Phi}_l \mathbf{D}_l) \ge r \}, \quad r > 0.
\end{align*}

\noindent where $g(m,x,l,\hat{q}) = \sum_{n=x}^{M} f(l,n) \zeta_x^{m,n}$, (b) follows Lemma \ref{le:C2}, i.e.,
\begin{align*}
& \sum_{x=r}^{m} g(m,x,l,q_i) > \sum_{x=r}^{m - y} g(m,x,l,q_i), \\
& r > 0, \; m > y > 0, q_i = q, \hat{q}
\end{align*}
and (c) is due to the assumption and Lemma \ref{le:C1}, i.e.,
\begin{align*}
& \sum_{x=r}^{m} g(m,x,l,\hat{q}) > \sum_{x=r}^{m} g(m,x,l,q),
\end{align*}

Therefore, the claim is proved by induction. With this claim, we then obtain
\begin{align*}
\hbar_{k+1} = \sum_{j = 0}^{M} j h_{k+1,j}
& = \sum_{j=1}^{M} \Pr \{ \mathbf{H}_{k+1} \ge j \} \\
& < \sum_{j=1}^{M} \Pr \{ \mathbf{\hat{H}}_{k+1} \ge j \} = \hat{\hbar}_{k+1}.
\end{align*}

The proof is completed.
\end{IEEEproof}

\begin{IEEEproof}[Proof of Theorem \ref{th:5}]
It is a direct consequence of Lemma \ref{le:C3} and the fact that the continuous relaxation of $t_i, i = 1, \ldots, l$, gives an extended feasible region containing the feasible region of $\mathbf{P1}$.
\end{IEEEproof}

\section{Proof of Proposition \ref{Prop:2}}\label{app-D}
\begin{lemma}
\label{le:D1}
\textit{
Let $\mathbf{\Phi} \in \mathbb{F}_q^{m \times t}$ and $ \mathbf{\hat{\Phi}} \in \mathbb{F}_{q}^{m \times \hat{t}}$ be totally random matrices, and let $\mathbf{D}$ and $ \mathbf{\hat{D}} $ be $t \times t$ and $\hat{t} \times \hat{t}$ random diagonal matrices, respectively. If $t < \hat{t}$, then $\Pr\{{\rm{rk}}(\mathbf{H} \mathbf{\Phi} \mathbf{D}) \ge r \} \le \Pr\{{\rm{rk}}(\mathbf{\hat{H}} \mathbf{\hat{\Phi}} \mathbf{\hat{D}}) \ge r \}$, $ 0 \le r \le k \le \min(M,m,t)$, where $\mathbf{H}$ and $\mathbf{\hat{H}}$ are arbitrary $M \times m$ matrices, and $\Pr \{ {\rm{rk}} (\mathbf{H}) \ge r \} \ge \Pr \{ {\rm{rk}} (\mathbf{\hat{H}}) \ge r \}$. The equality holds if and only if $r = 0$
}
\end{lemma}

\begin{IEEEproof}
Let $\mathbf{\hat{\Phi}}^{t}$ be a matrix consisting of arbitrary $t$ columns of $\mathbf{\hat{\Phi}}$, and let $X$ be the event that there are at least $x$ columns of $\mathbf{\hat{\Phi}}$ not belonging to space $ \langle \mathbf{\hat{\Phi}}^{t} \rangle$. We have
\begin{align*}
\Pr \{ \text{rk} (\mathbf{\hat{\Phi}}) \ge r \}
= & \Pr \{ \text{rk} (\mathbf{\hat{\Phi}}^{t}) \ge r \} \\
& + \sum_{i = 1}^{\min (\hat{t} - t, r)} \Pr \{ \text{rk} (\mathbf{\hat{\Phi}}^{t}) = r - i, X = i\}
\end{align*}

Since $\mathbf{\hat{\Phi}}^{t}$ is a $m \times t$ totally random matrix, we obtain
\begin{align*}
\Pr \{ \text{rk} (\mathbf{\hat{\Phi}}) \ge r \} \ge \Pr \{ \text{rk} (\mathbf{\hat{\Phi}}^t) \ge r \} = \Pr \{ \text{rk} (\mathbf{\Phi}) \ge r \},
\end{align*}

\noindent and then
\begin{align*}
\Pr \{ \text{rk} (\mathbf{\hat{H}} \mathbf{\hat{\Phi}}) \ge r \} \ge \Pr \{ \text{rk} (\mathbf{\hat{H}} \mathbf{\Phi}) \ge r \} \ge \Pr \{ \text{rk} (\mathbf{H} \mathbf{\Phi}) \ge r \}.
\end{align*}

Moreover, construct an $M \times \hat{t}$ matrix $\mathbf{A} = [\mathbf{H} \mathbf{\Phi} \quad \mathbf{0}]$, where $\mathbf{0}$ is an $M \times (\hat{t} - t)$ zero matrix. Since $\Pr \{ \text{rk} (\mathbf{H} \mathbf{\Phi}) \ge r \} = \Pr \{ \text{rk} (\mathbf{A}) \ge r \} \le \Pr \{ \text{rk} (\mathbf{\hat{\Phi}}) \ge r \}$ , we get
\begin{align*}
\Pr \{ \text{rk} (\mathbf{H} \mathbf{\hat{\Phi}} \mathbf{\hat{D}}) \ge r \} & \ge \Pr \{ \text{rk} (\mathbf{A} \mathbf{\hat{D}}) \ge r \} \\
& = \Pr \{ \text{rk} (\mathbf{H} \mathbf{\Phi} \mathbf{D}) \ge r \}.
\end{align*}

The proof is completed.
\end{IEEEproof}

\begin{IEEEproof}[Proof of Proposition \ref{Prop:2}]
Since $t_1^{\ast}$ is optimal, we have
\begin{align*}
\frac{\sum_{r=1}^{M} \alpha_r \beta_r g(r,t_1^{\ast})^{l_1}}{t_1^{\ast}} \ge \frac{\sum_{r=1}^{M} \alpha_r \beta_r g(r,t)^{l_1}}{t}.
\end{align*}

It can be verified that $g(r,t), r = 1, \ldots, M$, are monotonically increasing functions of $t$ by Lemma \ref{le:D1}. Consequently, for any $t \le t_1^{\ast}$, we obtain
\begin{align}
\label{eq:prop4.1}
\frac{\sum_{r=1}^{M} \alpha_r \beta_r g(r,t_1^{\ast})^{l_2}}{t_1^{\ast}} \ge \frac{\sum_{r=1}^{M} \alpha_r \beta_r g(r,t)^{l_2}}{t}.
\end{align}

Equation (\ref{eq:prop4.1}) shows that $t_2^{\ast}$ cannot be less than $t_1^{\ast}$.
\end{IEEEproof}


\begin{thebibliography}{99}

\bibitem{Fu05}
Z. Fu, H. Luo, P. Zerfos, S. Lu, L. Zhang, and M. Gerla, ``The impact of multihop wireless channel on TCP performance,'' \textit{IEEE Transactions on Mobile Computing}, 2005, 4(2), pp: 209-221.

\bibitem{Ahlswede00}
R. Ahlswede, N. Cai, S.-Y.R. Li, and R.W. Yeung, ``Network information flow,'' \textit{IEEE Transactions on Information Theory}, 2000, 46(4), pp: 1204-1206.

\bibitem{Li03}
S.-Y.R. Li,  R.W. Yeung, and  Ning Cai, ``Linear network coding,'' \textit{IEEE Transactions on Information Theory}, 2003, 49(2), pp: 371-381.

\bibitem{Chi10}
K. Chi, X. Jiang, and S. Horiguchi, ``Joint Design of Network Coding and Transmission Rate Selection for Multihop Wireless Networks,'' \textit{IEEE Transactions on Vehicular Technology}, 2010, 59(5), pp: 2435-2444.

\bibitem{Shokrollahi06}
A. Shokrollahi, ``Raptor codes,'' \textit{IEEE Transactions on Information Theory}, 2006, 52(6), pp: 2551-2567.

\bibitem{Luby11}
M. Luby, A. Shokrollahi, M. Watson, T. Stockhammer, and L. Minder,
“RaptorQ Forward Error Correction Scheme for Object Delivery,” Internet
Engineering Task Force (IETF), TS RFC6330, 2011. [Online]:
Available: http//:tools.ietf.org/html/rfc6330

\bibitem{YangTIT14}
S. Yang, and R. Yeung, ``Batched sparse codes,'' \textit{IEEE Transactions on Information Theory}, 60(9), 2014, pp. 5322-5346.

\bibitem{YangAllerton14}
S. Yang, R. W. Yeung, J. H.F. Cheung, and H. H.F. Yin, ``BATS: Network Coding in Action'', in Proc. 2014 52nd Annual Allerton Conference on in Communication, Control, and Computing (Allerton), Illinois, USA, Oct. 2014, pp: 1204-1211

\bibitem{YangLetter16}
S. Yang, and Q. Zhou, ``Tree Analysis of BATS Codes,'' \textit{IEEE Communications Letters}, 2016, 20(1), pp: 37-40.

\bibitem{Tang12}
B. Tang, S. Yang, Y. Yin, B. Ye, and S. Lu, ``Expander Graph Based Overlapped Chunked Codes'', in Proc. 2012 IEEE International Symposium on Information Theory Proceedings (ISIT), Cambridge, MA, USA, July 2012, pp: 2451-2455.

\bibitem{Tang16}
B. Tang, S. Yang, B. Ye, S. Guo, and S. Lu, ``Near-Optimal One-Sided Scheduling for Coded Segmented Network Coding,'' \textit{IEEE Transactions on Computers}, 2016, 65(3), pp: 929-939.

\bibitem{Mahdaviani12}
K. Mahdaviani, M. Ardakani, H. Bagheri, and C. Tellambura, ``Gamma Codes: A Low-Overhead Linear-Complexity Network Coding Solution'', in Proc. International Symposium on Network Coding (NetCod), 2012 International Symposium on, June 2012, pp: 125-130.

\bibitem{Ng13}
T. C. Ng and S. Yang, ``Finite-Length Analysis of Bats Codes,'' in Proc. Int.
Symp. Netw. Coding (NetCod), Calgary, AB, Canada, Jun. 2013, pp. 1-6.

\bibitem{Huang14}
Q. Huang, K. Sun, X. Li and D. Wu, ``Just FUN: A Joint Fountain Coding and Network Coding Approach to Loss-Tolerant Information Spreading,'' in Proc. The 15th ACM International Symposium on Mobile Ad Hoc Networking and Computing, Philadelphia USA, August 2014, pp: 83-92.

\bibitem{Zhang16}
H. Zhang, K. Sun, Q. Huang, Y. Wen, and D. Wu, ``FUN Coding: Design and Analysis,'' \textit{IEEE/ACM Transactions on Networking}, 2016, 24(6), pp: 3340-3353.

\bibitem{Xu16}
X. Xu, M. S.G.P. Kumar, Y.-L. Guan, and P. H.J. Chong, ``Two-Phase Cooperative Broadcasting Based on Batched Network Code,'' \textit{IEEE Transactions on Communications}, 2016, 64(2), pp: 706-714.

\bibitem{Yin16}
H.F. Yin, S. Yang, Q. Zhou, and L. M.L. Yung, ``Adaptive Recoding for BATS Codes,'' in Proc. 2016 IEEE International Symposium on Information Theory (ISIT), Barcelona, Spain, July 2016, pp: .

\bibitem{Muruganathan05}
S.D. Muruganathan, D.C.F. Ma, R.I. Bhasin, and  A.O. Fapojuwo, ``A Centralized Energy-Efficient Routing Protocol for Wireless Sensor Networks,'' \textit{IEEE Communications Magazine}, 2005, 43(3), pp: S8-13.

\bibitem{Li10}
Y. Li, Y. Jiang, D. Jin, L. Su, L. Zeng, and D. Wu, ``Energy-Efficient Optimal Opportunistic Forwarding for Delay-Tolerant Networks,'' \textit{IEEE Transactions on Vehicular Technology}, 2010,59(9), pp: 4500-4512.

\bibitem{Fulman15}
J. Fulman, and L. Goldstein, ``Stein’s Method and the Rank Distribution of Random Matrices over Finite Fields,'' \textit{The Annals of Probability}, 2015, 43(3), pp: 1274-1314.

\bibitem{Zelen72}
M. Zelen and N. C. Severo, ``Handbook of Mathematical Functions with Formulas, Graphs, and Mathematical Tables,'' M. Abramowitz and I. A. Stegun, Eds. New York: Dover, 1972.

\bibitem{Bassoli16}
R. Bassoli, V. N. Talooki, H. Marques, J. Rodriguez, R. Tafazolli, and S. Mumtaz, ``Hybrid Serial Concatenated Network Codes for Burst Erasure Channels,'' in Proc. 2015 IEEE 81st Vehicular Technology Conference (VTC Spring), Glasgow, UK, May 2015, pp: 1-4.

\bibitem{Zhao16}
H. Zhao, G. Dong, and H. Li, ``Simplified BATS Codes for Deep Space Multihop Networks,'' in Proc. IEEE Information Technology, Networking, Electronic and Automation Control Conference, Chongqing, China, May 2016, pp: 311-314.

\bibitem{La02}
R. J. La, and V. Anantharam, ``Utility-Based Rate Control in the Internet for Elastic Traffic,'' \textit{IEEE Transactions on Networking}, 2002, 10(2), pp: 272-286.

\bibitem{Perkins94}
C. E. Perkins, and P. Bhagwat, ``Highly Dynamic Destination Sequenced Distance-Vector Routing (DSDV) for Mobile Computers,'' in Proc. Conf. Commun. Archit., Protocols Appl., 1994, pp. 234–244.

\bibitem{Belotti09}
P. Belotti, J. Lee, L. Liberti, F. Margot, and A. Wächter, ``Branching and Bounds Tightening Techniques for Non-Convex MINLP,'' \textit{Optimization Methods and Software,} 2009, 24, pp: 597–634.

\bibitem{NOMAD}
S. Le Digabel, ``Algorithm 909: NOMAD: Nonlinear Optimization with the Mads Algorithm,'' \textit{ACM Transactions on Mathematical Software,} 2011, 37(4), PP: 1–15.

\bibitem{Gantovnik05}
V. B. Gantovnik, Z. Gurdal, L. T. Watson, and C. M. Anderson-Cook,
``Genetic Algorithm for Mixed Integer Nonlinear Programming Problems
using Separate Constraint Approximations,'' \textit(AIAA journal,) 2005, 43(8), pp. 1844–1849.

\bibitem{KNITRO}
KNITRO solver, https://www.artelys.com/en/optimization-tools/knitro.

\end{thebibliography}
\end{document}